\documentclass[12pt]{article}

\usepackage{geometry, graphicx, hyperref, cite, color, bm}
\usepackage{amsmath, amsfonts, amsthm, amssymb}
\numberwithin{equation}{section}

\begin{document}

\begin{titlepage}

\begin{center}
\vspace{1.0cm}
\Large{\textbf{PyCFTBoot: A flexible interface for the conformal bootstrap}}\\
\vspace{0.8cm}
\small{\textbf{Connor Behan}}\\
\vspace{0.5cm}
\textit{Department of Physics and Astronomy, Stony Brook University, Stony Brook, NY 11790, USA}\\
\textit{connor.behan@gmail.com}
\end{center}

\vspace{1.0cm}
\begin{abstract}
We introduce PyCFTBoot, a wrapper designed to reduce the barrier to entry in conformal bootstrap calculations that require semidefinite programming. Symengine and SDPB are used for the most intensive symbolic and numerical steps respectively. After reviewing the built-in algorithms for conformal blocks, we explain how to use the code through a number of examples that verify past results. As an application, we show that the multi-correlator bootstrap still appears to single out the Wilson-Fisher fixed points as special theories in dimensions between 3 and 4 despite the recent proof that they violate unitarity.
\end{abstract}

\end{titlepage}

\tableofcontents

\section{Introduction}
The conformal bootstrap \cite{fgg73, p74} has joined holography \cite{m98} as one of the most important tools for understanding strongly coupled conformal field theories (CFTs) in higher dimensions. Much of the progress comes from a numerical procedure initiated in \cite{rrtv08}, which exploits the constraints of crossing symmetry and unitarity. This has been successfully used to bound scaling dimensions and three point function coefficients in a wide range of conformal \cite{rv09, cr09, rrv10a, rrv10b, eppssv12, kps13, eppssv14, no14a, no14b, no14c, br14, cpy14} and superconformal \cite{ps10, psv11, brv13, clpy14, bllrv14} theories in dimensions between 2 and 6. The first widely released code designed to perform these calculations was \texttt{JuliBoots} \cite{p14}, a conformal bootstrap package based around a linear program solver. Shortly afterward, the solver \texttt{SDPB} \cite{s15} was released, giving the community access to the semidefinite programming methods pioneered in \cite{psv11, kps13, kps14}.\footnote{Readers interested in conformal blocks for their role in algebraic geometry might appreciate the \cite{s14} package.}

The advantages of the two are largely complementary. Semidefinite programming has superior performance in systems with multiple crossing equations and it is currently the only technique which extracts information from correlators of operators with different scaling dimensions \cite{kps14}. As such, \texttt{SDPB} has become the standard code for most numerical bootstrap studies in the last year \cite{kpsv15, cgikpy15, blrv15, ikppsy15a, ps15, ll15, lsswy15, cipy15, cp16}. Unlike \texttt{JuliBoots} however, it does not provide simple methods for specifying important kinematics information. Included in this are the crossing equations which depend on the type of CFT being studied and conformal blocks, special functions that depend on the dimension of space and a number of accuracy parameters. All of the above studies have performed these calculations using customized scripts for \texttt{Mathematica}. A new program, aiming to reduce this duplication of effort, is \texttt{PyCFTBoot} written in Python. Realizing a hope of \cite{p14}, it handles the computer algebra that goes into a numerical bootstrap entirely with free software. \texttt{PyCFTBoot} may be downloaded from
\begin{equation}
\texttt{https://github.com/cbehan/pycftboot} \nonumber
\end{equation}
where all future development is expected to take place. Besides \texttt{SDPB}, a few other dependencies are required in order to use it.

In mathematical Python software, \texttt{numpy} \cite{numpy} and \texttt{sympy} \cite{sympy} are two widely used packages that come to mind. Both of them are needed by \texttt{PyCFTBoot}. However, \texttt{sympy} is not fast enough to generate large tables of conformal blocks. It is only used in a few non-critical places that need to call Gegenbauer polynomials or the incomplete gamma function. Instead, the bulk of the symbolic algebra is handled by a fast C++ library called \texttt{symengine}. Python bindings have been chosen (over Ruby and Julia) because they are the most mature at the time of writing. These less common packages are downloadable from
\begin{eqnarray}
&& \texttt{https://github.com/symengine/symengine} \; \mathrm{(last \; tested: 5427bbe)} \; \nonumber \\
&& \texttt{https://github.com/symengine/symengine.py} \; \mathrm{(last \; tested: 9d23ef7)} \nonumber
\end{eqnarray}
Surprises are most easily avoided by using \texttt{PyCFTBoot} with Python 2.7 on GNU / Linux, but it has also been tested with Python 3.5. Descriptions of the important functions, included in the source code, may be viewed with the Python documentation server. Additionally, readers who are anxious to try the bootstrap may follow the commented tutorial distributed alongside the main file.

In section 2 of this note, we describe the algorithms that have been chosen to generate derivatives of conformal blocks and report some rough performance figures. Section 3 explains how semidefinite programs are formulated from these tables. In describing the main SDP object, it contains a few parts that read like passages from a user manual. Some examples, worked out in section 4, demonstrate that most of the known bootstrap results to date can in principle be reproduced with \texttt{PyCFTBoot}. Before we conclude, section 5 extends a previous result in the literature by using \texttt{PyCFTBoot} to probe the ``islands'' of allowed critical exponents in dimensions between 3 and 4 \cite{eppssv13}.

\section{Conformal Blocks}
Unlike with two or three point functions, conformal kinematics only determine the four point function up to an arbitrary dependence on two variables. Specifically for scalars,\footnote{We focus on the scalar correlators currently supported by \texttt{PyCFTBoot} but it would be very interesting to incorporate the ongoing work regarding operators with spin \cite{cppr11a, cppr11b, s12a, s12b, o12, d13, ch15, eks15, eeks15, ikppsy15a, rr15, ikppsy15b, eeks16}.}
\begin{equation}
\left < \phi_1(x_1) \phi_2(x_2) \phi_3(x_3) \phi_4(x_4) \right > = \left ( \frac{|x_{24}|}{|x_{14}|} \right )^{\Delta_{12}} \left ( \frac{|x_{14}|}{|x_{13}|} \right )^{\Delta_{34}} \frac{g(u, v)}{|x_{12}|^{\Delta_1 + \Delta_2}|x_{34}|^{\Delta_3 + \Delta_4}} \; , \label{4pt}
\end{equation}
where $u = \frac{x_{12}^2 x_{34}^2}{x_{13}^2 x_{24}^2}$ and $v = \frac{x_{14}^2 x_{23}^2}{x_{13}^2 x_{24}^2}$. As explained in the seminal works \cite{do00, do03} on (global) conformal blocks, $g(u, v)$ may be expanded in a convergent series with each term coming from a primary operator in the theory. This is done by way of the operator product expansion (OPE):
\begin{equation}
\phi_1(x) \phi_2(0) = \sum_{\mathcal{O}} \frac{\lambda_{12\mathcal{O}}}{|x|^{\Delta_1 + \Delta_2 - \Delta}} C_{\mathcal{O}}^{\mu_1 \dots \mu_\ell}(x, \partial) \mathcal{O}_{\mu_1 \dots \mu_\ell}(0) \; . \label{ope}
\end{equation}
Using this in the $(12)(34)$ channels for example produces $g(u, v) = \sum_{\mathcal{O}} \lambda_{12\mathcal{O}} \lambda_{34\mathcal{O}} g_{\mathcal{O}}^{\Delta_{12}, \Delta_{34}}(u, v)$ where each function depends on the spatial dimension $d$ or equivalently on $\nu = \frac{d - 2}{2}$. The subscript $\mathcal{O}$ is often written as $(\Delta, \ell)$ since all primary operators that couple to scalars transform in some spin-$\ell$ representation of $SO(d)$. Crossing symmetry is the statement that all three choices for the OPE channels must agree. This is what leads to the bootstrap but we will postpone a discussion of this to the next section.

\subsection{Rational approximations}
Rather than the cross-ratios $u$ and $v$, conformal blocks are most often considered as functions of $z$ and $\bar{z}$, defined by using conformal transformations to send $x_1$, $x_3$ and $x_4$ to 0, 1 and $\infty$ respectively. The blocks are analytic for $0 < z, \bar{z} < 1$ and most bootstrap studies focus on the crossing symmetric point $(z_*, \bar{z}_*) = (\frac{1}{2}, \frac{1}{2})$. Although there are other useful variables \cite{fkp13}, Table \ref{coords} shows all of the co-ordinates used by \texttt{PyCFTBoot}.
\begin{table}
\begin{center}
\begin{tabular}{l|l|l}
First & Second & Crossing point \\
\hline
$u = |z|^2$ & $v = |1 - z|^2$ & $(u_*, v_*) = \left ( \frac{1}{4}, \frac{1}{4} \right )$ \\
$a = z + \bar{z}$ & $b = (z - \bar{z})^2$ & $(a_*, b_*) = (1, 0)$ \\
$\rho = \frac{z}{(1 + \sqrt{1 - z})^2}$ & $\bar{\rho} = \frac{\bar{z}}{(1 + \sqrt{1 - \bar{z}})^2}$ & $(\rho_*, \bar{\rho}_*) = (3 - 2\sqrt{2}, 3 - 2\sqrt{2})$ \\
$r = |\rho|$ & $\eta = \frac{\rho + \bar{\rho}}{2|\rho|}$ & $(r_*, \eta_*) = (3 - 2\sqrt{2}, 1)$
\end{tabular}
\end{center}
\vspace{-0.5cm}
\caption{Useful variables for four point conformal blocks in terms of $z$ and $\bar{z}$.}
\label{coords}
\end{table}
As observed in \cite{hr13, kps13}, a block may be expanded in powers of $r$ where each term corresponds to a new descendant in the multiplet of $\mathcal{O}$. As the scaling dimension $\Delta$ is varried, coefficients in the sum diverge at certain non-unitary values. When they do, the residue is proportional to a conformal block itself. This motivated \cite{kps13} to develop the recurrence relations
\begin{eqnarray}
h_{\Delta, \ell}^{\Delta_{12}, \Delta_{34}}(r, \eta) &\equiv& r^{-\Delta} g_{\Delta, \ell}^{\Delta_{12}, \Delta_{34}}(r, \eta) \nonumber \\
h_{\Delta, \ell}^{\Delta_{12}, \Delta_{34}}(r, \eta) &=& h_{\infty, \ell}^{\Delta_{12}, \Delta_{34}}(r, \eta) + \sum_i \frac{c_i^{\Delta_{12}, \Delta_{34}}(\ell) r^{n_i}}{\Delta - \Delta_i(\ell)} h_{\Delta_i(\ell) + n_i, \ell_i}^{\Delta_{12}, \Delta_{34}}(r, \eta) \; . \label{recurse1}
\end{eqnarray}
The leading term is given by \cite{kps14}
\begin{equation}
h_{\infty, \ell}^{\Delta_{12}, \Delta_{34}}(r, \eta) = \frac{\ell!}{(2\nu)_{\ell}} \frac{(-1)^{\ell} C_{\ell}^{\nu}(\eta)}{(1 - r^2)^{\nu}(1 + r^2 + 2r\eta)^{\frac{1}{2}(1 + \Delta_{12} - \Delta_{34})}(1 + r^2 - 2r\eta)^{\frac{1}{2}(1 - \Delta_{12} + \Delta_{34})}} \; . \label{leading-block}
\end{equation}
\begin{table}
\begin{center}
\begin{tabular}{c|c|c|c}
$n_i$ & $\Delta_i(\ell)$ & $\ell_i$ & $c_i^{\Delta_{12}, \Delta_{34}}(\ell)$ \\
\hline
$k$ & $1 - \ell - k$ & $\ell + k$ & $c_1^{\Delta_{12}, \Delta_{34}}(\ell, k)$ \\
$2k$ & $1 + \nu - k$ & $\ell$ & $c_2^{\Delta_{12}, \Delta_{34}}(\ell, k)$ \\
$k$ & $1 + \ell + 2\nu - k$ & $\ell - k$ & $c_3^{\Delta_{12}, \Delta_{34}}(\ell, k)$
\end{tabular}
\end{center}
\vspace{-0.5cm}
\caption{The three types of poles in $\Delta$ for the meromorphic conformal blocks. Two of them have infinitely many elements labelled by the integer $k > 0$. The third type requires $0 < k \leq \ell$.}
\label{data}
\end{table}
Table \ref{data} describes the data needed to construct the poles and residues in (\ref{recurse1}). These were noticed empirically in \cite{kps14} but most of them were later proven in \cite{pty15}. We must use
\begin{eqnarray}
c_1^{\Delta_{12}, \Delta_{34}}(\ell, k) &=& -\frac{k(-4)^k}{(k!)^2} \frac{(\ell + 2\nu)_k}{(\ell + \nu)_k} \left ( \frac{1}{2} (1 - k + \Delta_{12}) \right )_k \left ( \frac{1}{2} (1 - k + \Delta_{34}) \right )_k \nonumber \\
c_2^{\Delta_{12}, \Delta_{34}}(\ell, k) &=& \frac{k (\nu + 1)_{k - 1} (-\nu)_{k + 1}}{(k!)^2} \frac{\ell + \nu - k}{\ell + \nu + k} \left ( \frac{\ell + \nu - k + 1}{2} \right )_k^{-2} \left ( \frac{\ell + \nu - k}{2} \right )_k^{-2} \nonumber \\
&& \left ( \frac{1}{2} (1 - k + \ell - \Delta_{12} + \nu) \right )_k \left ( \frac{1}{2} (1 - k + \ell + \Delta_{12} + \nu) \right )_k \nonumber \\
&& \left ( \frac{1}{2} (1 - k + \ell - \Delta_{34} + \nu) \right )_k \left ( \frac{1}{2} (1 - k + \ell + \Delta_{34} + \nu) \right )_k \; \label{residues} \\
c_3^{\Delta_{12}, \Delta_{34}}(\ell, k) &=& -\frac{k(-4)^k}{(k!)^2} \frac{(\ell + 1 - k)_k}{(\ell + \nu + 1 - k)_k} \left ( \frac{1}{2} (1 - k + \Delta_{12}) \right )_k \left ( \frac{1}{2} (1 - k + \Delta_{34}) \right )_k \nonumber
\end{eqnarray}
to fill in the last column. One fact that can be seen from (\ref{residues}) is that $c_1^{0,0}(\ell, k)$ and $c_3^{0,0}(\ell, k)$ are only non-zero when $k$ is even. This means that when the external scalars are identical, blocks of even and odd spin do not show up in each other's recurrence relations. Consequently, adjusting the overall normalization of $h_{\Delta, \ell}^{0, 0}(r, \eta)$ by $(-1)^{\ell}$ is equivalent to simply removing the factor of $(-1)^{\ell}$ from (\ref{leading-block}). Indeed, for many studies involving identical scalars, it was not present. The generalization to non-zero dimension differences shows us that more drastic changes would be needed if we still wanted to cancel the $(-1)^{\ell}$ in (\ref{leading-block}). Therefore \texttt{PyCFTBoot} keeps it around. The end of this paper points out the examples in which this subtlety needs to be remembered.

For spins up to some $\ell_{\mathrm{max}}$, we need to know several derivatives of $h_{\Delta, \ell}^{\Delta_{12}, \Delta_{34}}$ evaluated at $(r_*, \eta_*) = (3 - 2\sqrt{2}, 1)$. If we evaluated (\ref{recurse1}) for powers of $r$ up to $k_{\mathrm{max}}$ and differentiated after, we would suffer a large performance hit. This is because there would be many appearances of (\ref{leading-block})'s non-polynomial contributions all multiplied by different powers of $r$. A better strategy is to compute all derivatives at the same time via matrix multiplication \cite{eppssv14}. To this end, we define the vector $\textbf{h}_{\infty, \ell}$ with all desired derivatives of (\ref{leading-block}) already evaluated at the crossing point. They are grouped into ``chunks'' of $\partial_r$ powers for a given number of $\partial_\eta$ powers.\footnote{Although we describe the general case here, we will soon see that normal use of \texttt{PyCFTBoot} will only involve one chunk.} For example, a computation going up to third order would set
\begin{equation}
\textbf{h}_{\infty, \ell} = \left [ 1 \; \frac{\partial}{\partial r} \; \frac{\partial^2}{\partial r^2} \; \frac{\partial^3}{\partial r^3} \; \frac{\partial}{\partial \eta} \; \frac{\partial^2}{\partial \eta \partial r} \; \frac{\partial^3}{\partial \eta \partial r^2} \; \frac{\partial^2}{\partial \eta^2} \; \frac{\partial^3}{\partial \eta^2 \partial r} \; \frac{\partial^3}{\partial \eta^3} \right ]^{\mathrm{T}} h_{\infty, \ell} \; .
\end{equation}
Seeing what happens when we differentiate $r^{n_i} h_{\Delta, \ell}$ several times, the matrix telling us what linear combination of derivatives to take is
\begin{equation}
\textbf{R}^{n_i} = \left [
\begin{tabular}{llll}
$r_*^{n_i}$ & $0$ & $0$ & \dots \\
$n_i r_*^{n_i - 1}$ & $r_*^{n_i}$ & $0$ & \dots \\
$n_i (n_i - 1) r_*^{n_i - 2}$ & $2n_i r_*^{n_i - 1}$ & $r_*^{n_i}$ & \dots \\
$\vdots$ & $\vdots$ & $\vdots$ & $\ddots$
\end{tabular}
\right ] = \left [
\begin{tabular}{llll}
$r_*$ & $0$ & $0$ & \dots \\
$1$ & $r_*$ & $0$ & \dots \\
$0$ & $2$ & $r_*$ & \dots \\
$\vdots$ & $\vdots$ & $\vdots$ & $\ddots$
\end{tabular}
\right ]^{n_i} \; . \label{r-matrix}
\end{equation}
This is the matrix acting on a single chunk. Since $\eta$ is unaffected, the full $\textbf{R}$ is the tensor product of (\ref{r-matrix}) with the identity. There is a problem with simply writing
\begin{equation}
\textbf{h}_{\Delta, \ell} = \textbf{h}_{\infty, \ell} + \sum_i \frac{c_i(\ell) \textbf{R}^{n_i}}{\Delta - \Delta_i(\ell)} \textbf{h}_{\Delta_i(\ell) + n_i, \ell_i}
\end{equation}
and repeating this calculation every time a new block appears. It is most easily seen if we compare the number of matrix multiplications involved to the number of unique $\textbf{h}_{\Delta_i + n_i, \ell_i}$ terms introduced by the recursion. Looking at (\ref{residues}), we see a residue $c_2(\ell, k)$ which may vanish sometimes and a residue $c_3(\ell, k)$ which only exists for certain spins. Therefore, the best case scenario (only using $c_1(\ell, k)$) tells us that the number of matrix multiplications $\#$ satisfies
\begin{eqnarray}
\#(0) &=& 1 \nonumber \\
\#(k_{\mathrm{max}}) &>& \sum_{k = 0}^{k_{\mathrm{max}} - 1} \#(k) \; .
\end{eqnarray}
This is the same relation satisfied by the partition function which counts the number of ways to write an integer as the sum of smaller ones. The well known asymptotics of this function \cite{dmmv00}, tell us that duplicated matrix multiplications will abound by many orders of magnitude with this naive method. Instead \texttt{PyCFTBoot} again follows \cite{eppssv14} and predicts which residues will be needed ahead of time. This is simply a matter of letting the spin take values $\ell \leq \ell_{\mathrm{max}} + k_{\mathrm{max}}$ for a table whose final entires describe spins up to $\ell_{\mathrm{max}}$. For each value of $\ell$, we let the index $i$ run over all admissible poles in Table \ref{data} and define the residue vectors $\textbf{d}_{\ell, i}$. All of these are initialized to $\textbf{h}_{\infty, \ell_i}$. It is then straightforward to iterate
\begin{equation}
\textbf{d}_{\ell, i} = c_i(\ell) \textbf{R}^{n_i} \left [ \textbf{h}_{\infty, \ell_i} + \sum_j \frac{\textbf{d}_{\ell_i, j}}{\Delta_i(\ell) + n_i - \Delta_j(\ell_i)} \right ] \label{recurse2}
\end{equation}
and stop once enough powers of $\textbf{R}$ are introduced. Rather than updating the residues right away, we consider all $\textbf{d}_{\ell, i}$ on the right hand side to be the ``old values'' and replace them with the ``new values'' once everything on the left hand side has been calculated. These go into the expression
\begin{equation}
\textbf{h}_{\Delta, \ell} = \textbf{h}_{\infty, \ell} + \sum_i \frac{\textbf{d}_{\ell, i}}{\Delta - \Delta_i(\ell)} \; . \label{meromorphic-block}
\end{equation}
It is clear that the entries in $\textbf{h}_{\Delta, \ell}$ are rational functions of $\Delta$. They all have different numerators and the same denominator. Instead of computing (\ref{meromorphic-block}) as written and taking extra time to extract the numerator and denominator, \texttt{PyCFTBoot} stores them separately from the start. The leading term of (\ref{meromorphic-block}) is multiplied by $\prod_j (\Delta - \Delta_j(\ell))$ and the $i^{\mathrm{th}}$ term of it is multiplied by $\prod_{j \neq i} (\Delta - \Delta_j(\ell))$.

There is a modification to (\ref{meromorphic-block}) that can be used to produce polynomials of smaller degree. Described in \cite{kps13}, it slightly increases the time needed to generate a conformal block table but it can greatly decrease the running time of \texttt{SDPB}. The idea is to split the set of poles $\mathcal{P}$ into ``large and small'' types and use the poles of $\mathcal{P}_>$ to approximate those in $\mathcal{P}_<$. As our crieterion, we check whether the maximum component of $\textbf{d}_{\ell, i}$ is above or below some cutoff $\theta$. For $\Delta_i \in \mathcal{P}_<$, we attempt to choose the $a_{i, k}$ coefficients optimally in
\begin{equation}
\frac{1}{\Delta - \Delta_i} \approx \sum_{\Delta_k \in \mathcal{P}_>} \frac{a_{i, k}}{\Delta - \Delta_k} \; . \label{pole-approximation}
\end{equation}
Following the choice in \cite{kps13}, we demand that the first $|\mathcal{P}_>| / 2$ derivatives of (\ref{pole-approximation}) hold exactly at $\Delta = \Delta_{\mathrm{unitary}} + \theta$ and $\Delta = \theta^{-1}$. If $|\mathcal{P}_>|$ is odd, the last of these derivatives will only hold at one of the points. Once the $a_{i, k}$ are determined by this invertible linear system, \texttt{PyCFTBoot} incorporates them into the calculation of (\ref{meromorphic-block}). Whenever it needs to multiply by $\prod_{\Delta_j \neq \Delta_i} (\Delta - \Delta_j)$ and $\Delta_i \in \mathcal{P}_<$, it instead multiplies by $\sum_{\Delta_k \in \mathcal{P}_>} a_{i, k} \prod_{\Delta_j \in \mathcal{P}_> \setminus \{ \Delta_k \} } (\Delta - \Delta_j)$.

After the (\ref{meromorphic-block}) computation with the optional degree reduction step, one must obtain a vector $\textbf{g}_{\Delta, \ell}$ of true conformal block derivatives from its meromorphic version $\textbf{h}_{\Delta, \ell}$. This is done by restoring the $r_*^{\Delta}$ singularity with another matrix. Specifically,
\begin{equation}
\textbf{g}_{\Delta, \ell} = r_*^{\Delta} \textbf{S} \textbf{h}_{\Delta, \ell} \; . \label{conformal-block}
\end{equation}
It is easy to see that $r_*^{\Delta} \textbf{S}$ must be the same matrix as $\textbf{R}^{n_i}$ in (\ref{r-matrix}) with $n_i$ replaced by $\Delta$. There is no need to build up $\textbf{S}$ by repeatedly multiplying some simpler matrix by itself. Its $(i, j)$ element is immediately known to be $\frac{\Delta \dots (\Delta - j)}{r_*^j} \binom{i}{j}$. Elements of the conformal block vector continue to be rational functions. However, if all numerators in $\textbf{h}_{\Delta, \ell}$ have the same degree, those in $\textbf{g}_{\Delta, \ell}$ will have a degree that increases with the order of the derivative. Looking at these numerators, the end result is something of the form
\begin{equation}
\frac{\partial^{m + n}}{\partial \eta^m \partial r^n} g_{\Delta, \ell}^{\Delta_{12}, \Delta_{34}}(r_*, \eta_*) = \chi_{\ell}(\Delta) P_{\ell}^{\Delta_{12}, \Delta_{34} ; mn}(\Delta) \label{pos-poly1}
\end{equation}
which is a polynomial times the positive function $\chi_{\ell}(\Delta) = r_*^{\Delta} \prod_j (\Delta - \Delta_j(\ell))^{-1}$. This is precisely the form required for a task that involves semidefinite programming.

\subsection{Further processing}
Going from the $(12)(34)$ to the $(14)(23)$ channel switches $u \leftrightarrow v$ and modifies the prefactor in the four point function (\ref{4pt}). Crossing equations are obtained by setting the differences of these four point functions to zero. The simplest crossing equation with no global symmetry is $v^{\frac{\Delta_2 + \Delta_3}{2}} g_{1234}(u, v) - u^{\frac{\Delta_1 + \Delta_2}{2}} g_{3214}(v, u) = 0$ \cite{kps14}. As a result, functions of the form
\begin{equation}
F_{\pm, \Delta, \ell}(u, v) = v^{\Delta_{\phi}} g_{\Delta, \ell}^{\Delta_{12}, \Delta_{34}}(u, v) \pm u^{\Delta_{\phi}} g_{\Delta, \ell}^{\Delta_{12}, \Delta_{34}}(v, u) \; , \label{convolved-block-schematic}
\end{equation}
are the natural objects to consider once conformal blocks are known. These have come to be called convolved conformal blocks \cite{p14}. In principle, convolved conformal blocks and their derivatives could be calculated directly from the (\ref{pos-poly1}) result with its $r$ and $\eta$ variables. However, the simple $u \leftrightarrow v$ transformation is represented by $r$ and $\eta$ in a much more complicated way. When the second half of (\ref{convolved-block-schematic}) involves a new function $g_{\Delta, \ell}^{\Delta_{12}, \Delta_{34}}(\tilde{r}(r, \eta), \tilde{\eta}(r, \eta))$, much of the work that goes into the $\frac{\partial^{m + n}}{\partial \eta^m \partial r^n} F_{\pm, \Delta, \ell}(r_*, \eta_*)$ calculation will be spent differentiating $\tilde{r}$ and $\tilde{\eta}$. This extra work during the convolution step can be eliminated if we instead add extra work during the conformal block step to convert (\ref{pos-poly1}) to $(z, \bar{z})$ or $(a, b)$ variables. At first glance, it might seem that the benefit of this choice is purely organizational --- it allows the fast and slow calculations in \texttt{PyCFTBoot} to be conceptually separate. As we now discuss however, there is another recurrence relation which gives us a much stronger incentive to change variables.

Conformal blocks are eigenfunctions of the quadratic Casimir \cite{do03}:
\begin{equation}
\left [ D_z + D_{\bar{z}} + 2\nu \frac{z \bar{z}}{z - \bar{z}} \left ( (1 - z) \frac{\textup{d}}{\textup{d}z} - (1 - \bar{z}) \frac{\textup{d}}{\textup{d}\bar{z}} \right ) \right ] g_{\Delta, \ell}^{\Delta_{12}, \Delta_{34}} = c_2 g_{\Delta, \ell}^{\Delta_{12}, \Delta_{34}} \; . \label{casimir}
\end{equation}
Here, the definitions
\begin{eqnarray}
D_z &=& (1 - z)z^2 \frac{\textup{d}^2}{\textup{d}z^2} + \left ( \frac{1}{2} \Delta_{12} - \frac{1}{2} \Delta_{34} - 1 \right ) \frac{\textup{d}}{\textup{d}z} + \frac{1}{4} \Delta_{12} \Delta_{34} z \nonumber \\
c_2 &=& \frac{1}{2} \left [ \ell (\ell + 2\nu) + \Delta(\Delta - 2 - 2\nu) \right ] \; ,
\end{eqnarray}
are standard. The existence of a linear differential equation satisfied by the blocks suggests the possibility of building up high order derivatives from lower ones. We may pretend for a minute that $g_{\Delta, \ell}$, $\frac{\partial g_{\Delta, \ell}}{\partial z}$ and $\frac{\partial^2 g_{\Delta, \ell}}{\partial z \partial \bar{z}}$ are all known at $\left ( \frac{1}{2}, \frac{1}{2} \right )$. The content of (\ref{casimir}) is then to tell us what $\frac{\partial^2 g_{\Delta, \ell}}{\partial z^2}$ is at the same point. We could attempt to continue this pattern by differentiating (\ref{casimir}) with respect to $z$ but then $\frac{\partial^3 g_{\Delta, \ell}}{\partial z^3}$ would not be the only unknown derivative anymore. The presence of new unknowns like $\frac{\partial^3 g_{\Delta, \ell}}{\partial \bar{z}^2 \partial z}$ forces us to use something more clever.

Such cleverness was found by \cite{hor13} in which the quadratic and quartic Casimirs of the conformal group are used together. This reveals an ordinary differential equation satisfied by the blocks on the $z = \bar{z}$ diagonal. In terms of the $a$ co-ordinate, this new equation (which clearly keeps new derivatives under control) is
\begin{eqnarray}
&& D_a^{(4, 3)} g_{\Delta, \ell}^{\Delta_{12}, \Delta_{34}} = 0 \; \label{recurse3} \\
&& D_a^{(4, 3)} \equiv \left ( \frac{a}{2} - 1 \right )^3 a^4 \frac{\textup{d}^4}{\textup{d}a^4} + p_3 \left ( \frac{a}{2} - 1 \right )^2 a^3 \frac{\textup{d}^3}{\textup{d}a^3} + p_2 \left ( \frac{a}{2} - 1 \right ) a^2 \frac{\textup{d}^2}{\textup{d}a^2} + p_1 a \frac{\textup{d}}{\textup{d}a} + p_0 \; . \nonumber
\end{eqnarray}
The polynomials $p_0, \dots, p_3$ used by \texttt{PyCFTBoot} are the ones in \cite{hor13} except with a slight change: they are written with $\frac{a}{2}$ in place of $z$ and multiplied by 8 to force as many coefficients as possible to still be integers. Differentiating (\ref{recurse3}), a fifth derivative of $g_{\Delta, \ell}^{\Delta_{12}, \Delta_{34}}$ becomes the highest order term. However, the lowest order term continues to be a zeroth derivative. Because $p_0(a)$ has degree 3, our equation only stops having non-derivative terms once it goes up to $\frac{\textup{d}^8 g_{\Delta, \ell}}{\textup{d}a^8}$. This means that the $m^{\mathrm{th}}$ diagonal derivative is calculated from the $\mathrm{min}(m, 7)$ lower ones using a handful of simple polynomials. One only needs $m$ to be at least 4 in order to start this process. Because of this, vectors in the slow original recursion (\ref{recurse2}) only need to fit four $\partial_r$ powers. Once the $a$ derivatives are known, more recurrence relations determine the $b$ derivatives. Defining $S = -\frac{1}{2} \left ( \Delta_{12} - \Delta_{34} \right )$ and $P = -\frac{1}{2} \Delta_{12} \Delta_{34}$, we use
\begin{eqnarray}
&& 2(1 - 2n - 2\nu) \frac{\partial^{m + n} g_{\Delta, \ell}}{\partial a^m \partial b^n} = \nonumber \\
&& 2m(1 - 2n - 2\nu) \left [ -\frac{\partial^{m + n - 1} g_{\Delta, \ell}}{\partial a^{m - 1} \partial b^n} + (m - 1) \frac{\partial^{m + n - 2} g_{\Delta, \ell}}{\partial a^{m - 2} \partial b^n} + (m - 1)(m - 2) \frac{\partial^{m + n - 3} g_{\Delta, \ell}}{\partial a^{m - 3} \partial b^n} \right ] \nonumber \\
&& + \frac{\partial^{m + n + 1} g_{\Delta, \ell}}{\partial a^{m + 2} \partial b^{n - 1}} - (6 - m - 4n + 2\nu + 2S) \frac{\partial^{m + n} g_{\Delta, \ell}}{\partial a^{m + 1} \partial b^{n - 1}} \nonumber \\
&& - \left [ 4c_2 + m^2 + 8mn - 5m + 4n^2 - 2n - 2 \right . \nonumber \\
&& \left . - 4\nu(1 - m - n) + 4S(m + 2n - 2) + 2P \right ] \frac{\partial^{m + n - 1} g_{\Delta, \ell}}{\partial a^m \partial b^{n - 1}} \nonumber \\
&& - m \left [ m^2 + 12mn - 13m + 12n^2 - 34n + 22 \right . \nonumber \\
&& \left . - 2\nu(2n - m - 1) + 2S(m + 4n - 5) + 2P \right ] \frac{\partial^{m + n - 2} g_{\Delta, \ell}}{\partial a^{m - 1} \partial b^{n - 1}} \nonumber \\
&& + (1 - n) \left [ \frac{\partial^{m + n} g_{\Delta, \ell}}{\partial a^{m + 2} \partial b^{n - 2}} - (6 - 3m - 4n + 2\nu - 2S) \frac{\partial^{m + n - 1} g_{\Delta, \ell}}{\partial a^{m + 1} \partial b^{n - 2}} \right ] \; . \label{recurse4}
\end{eqnarray}
This is the transverse derivative recursion found in \cite{eppssv12} generalized to unequal external dimensions with the different definition of $c_2$ taken into account. It follows from going back to the original Casimir PDE (\ref{casimir}) in the $(a, b)$ co-ordinates. The same coefficients can also be found in recent versions of the \cite{p14} source code. The form of (\ref{recurse4}) tells us the shape that will be taken by a lattice of derivatives we compute this way. When we make $m$ as high as possible for a given $n$, the right hand side shows that 2 must be added to reach the highest possible $m$ for $n - 1$. This leads to the triangle
\begin{eqnarray}
n &\in& \{0, \dots, n_{\mathrm{max}} \} \nonumber \\
m &\in& \{0, \dots, 2(n_{\mathrm{max}} - n) + m_{\mathrm{max}} \} \; , \label{triangle}
\end{eqnarray}
depending on two user-defined parameters. As found in \cite{ep13}, a high $n_{\mathrm{max}}$ is more important than a high $m_{\mathrm{max}}$. An obvious point worth remembering is that (\ref{recurse3}) and (\ref{recurse4}) are only satisfied by exact conformal blocks, not their rational approximations. As a result, these recursions are only valid for computing derivatives if $k_{\mathrm{max}}$ is sufficiently large. 

Returning to the task of convolution, we need to compute derivatives of
\begin{equation}
F_{\pm, \Delta, \ell}(a, b) = \left ( \frac{(2 - a)^2 - b}{4} \right )^{\Delta_{\phi}} g_{\Delta, \ell}^{\Delta_{12}, \Delta_{34}}(a, b) \pm \left ( \frac{a^2 - b}{4} \right )^{\Delta_{\phi}} g_{\Delta, \ell}^{\Delta_{12}, \Delta_{34}}(2 - a, b) \; , \label{convolved-block}
\end{equation}
at $(a_*, b_*) = (1, 0)$. We may immediately see that only one of the two terms in (\ref{convolved-block}) needs to be differentiated. If the number of $a$ derivatives is even (odd), the other term will contribute equally (oppositely) for $F_{+, \Delta, \ell}$ and oppositely (equally) for $F_{-, \Delta, \ell}$. We therefore reduce one vector of derivatives to another vector of derivatives having roughly half the size. As in the unconvolved case, its components have the positive-times polynomial form. Knowing that $\Delta_\phi$ will eventually be determined by the external dimensions $\Delta_i, \Delta_j, \Delta_k, \Delta_l$, we write
\begin{equation}
\frac{\partial^{m + n}}{\partial a^m \partial b^n} F_{\pm, \Delta, \ell}^{ij ; kl}(a_*, b_*) = \chi_{\ell}(\Delta) P_{\pm, \ell}^{ij ; kl ; mn}(\Delta) \; . \label{pos-poly2}
\end{equation}
The linear combinations we need to take in order to compute these polynomials are known in closed form. For the following calculation, it is easiest to take all of the $b$ derivatives first and then set $b = 0$. This allows us to treat all terms as being linear in $a$.
\begin{eqnarray}
\frac{\partial^{m + n}}{\partial a^m \partial b^n} \left ( \frac{(2 - a)^2 - b}{4} \right )^{\Delta_{\phi}} g_{\Delta, \ell} &=& \sum_{i = 0}^m \sum_{j = 0}^n \binom{m}{i} \binom{n}{j} \frac{\partial^{i + j}}{\partial a^i \partial b^j} \left ( \frac{(2 - a)^2 - b}{4} \right )^{\Delta_{\phi}} \frac{\partial^{m + n - i - j} g_{\Delta, \ell}}{\partial a^{m - i} \partial b^{n - j}} \nonumber \\
&\rightarrow& \sum_{i = 0}^m \sum_{j = 0}^n \binom{m}{i} \binom{n}{j} \left ( \frac{1}{4} \right )^j \left ( -\Delta_{\phi} \right )_j \nonumber \\
&& \frac{\partial^i}{\partial a^i} \left ( 1 - \frac{a}{2} \right )^{2\Delta_{\phi} - 2j} \frac{\partial^{m + n - i - j} g_{\Delta, \ell}}{\partial a^{m - i} \partial b^{n - j}} \nonumber \\
&=& \sum_{i = 0}^m \sum_{j = 0}^n \binom{m}{i} \binom{n}{j} \left ( \frac{1}{4} \right )^j \left ( \frac{1}{2} \right )^i \left ( -\Delta_{\phi} \right )_j \left ( 2j - 2\Delta_{\phi} \right )_i \label{convolution} \\
&& \left ( 1 - \frac{a}{2} \right )^{2\Delta_{\phi} - 2j - i} \frac{\partial^{m + n - i - j} g_{\Delta, \ell}}{\partial a^{m - i} \partial b^{n - j}} \nonumber \\
&\rightarrow& \sum_{i = 0}^m \sum_{j = 0}^n \binom{m}{i} \binom{n}{j} \left ( \frac{1}{4} \right )^{\Delta_{\phi}} \left ( -\Delta_{\phi} \right )_j \left ( 2j - 2\Delta_{\phi} \right )_i \frac{\partial^{m + n - i - j} g_{\Delta, \ell}}{\partial a^{m - i} \partial b^{n - j}} \nonumber
\end{eqnarray}
We may now summarize how the input parameters $d, k_{\mathrm{max}}, \ell_{\mathrm{max}}, m_{\mathrm{max}}, n_{\mathrm{max}}, \Delta_{12}, \Delta_{34}$ are used to prepare a conformal bootstrap environment. \texttt{PyCFTBoot},
\begin{enumerate}
\item Creates a vector $\textbf{h}_{\infty, \ell}$ containing $r$ derivatives of (\ref{leading-block}) up to third order.
\item Calculates $\textbf{d}_{\ell, i}$ residues with $k_{\mathrm{max}}$ iterations that use the data in (\ref{residues}) and Table \ref{data}.
\item Combines these into $\ell_{\mathrm{max}}$ meromorphic blocks $\textbf{h}_{\Delta, \ell}$ through (\ref{meromorphic-block}), optionally approximating small poles with (\ref{pole-approximation}).
\item Converts these into genuine conformal blocks $\textbf{g}_{\Delta, \ell}$ with the matrix (\ref{conformal-block}).
\item Applies the chain rule to get $\mathrm{min}(m_{\mathrm{max}} + 2n_{\mathrm{max}}, 3)$ derivatives of all $g_{\Delta, \ell}^{\Delta_{12}, \Delta_{34}}$ with respect to $a$.
\item Uses (\ref{recurse3}), (\ref{recurse4}) to calculate whatever $a$, $b$ derivatives are left and then uses (\ref{convolution}) to take the convolution leaving $\Delta_{\phi}$ as a free variable.
\end{enumerate}
Almost all of the time is spent on the first four steps. The most interesting parameter here is $k_{\mathrm{max}}$ because these steps clearly have a running time which is sublinear in the number of spins. Table \ref{stats} times the calculation of a few increasingly accurate conformal block tables with the single spin $\ell = 0$.
\begin{table}
\begin{center}
\begin{tabular}{c|ccc}
$k_{\mathrm{max}}$ & Trial 1 & Trial 2 & Trial 3 \\
\hline
10 & 1.106 & 1.119 & 1.111 \\
15 & 3.146 & 3.151 & 3.127 \\
20 & 6.859 & 6.856 & 6.873 \\
25 & 13.731 & 13.844 & 13.682 \\
30 & 23.062 & 23.058 & 23.131
\end{tabular}
\end{center}
\vspace{-0.5cm}
\caption{Running time in seconds for the $d = 3$ calculation of $g_{\Delta, 0}^{0, 0}(a, b)$ and its first three derivatives with respect to $a$. Three trials were done on one core of a 2.4GHz machine.}
\label{stats}
\end{table}

\subsection{Even dimensions}
Unlike the exact expressions for conformal blocks \cite{do00, do03, do11} which are only known in even dimension, the scheme based on (\ref{recurse2}) that we have described so far works best when $d$ is odd or fractional. This is because it assumes that all poles in $\Delta$ for a conformal block are simple. The breakdown of this assumption as $\nu$ becomes an integer can be seen as certain poles approach each other and certain residues diverge. In order to handle even dimensions, \texttt{PyCFTBoot} switches to a different method for calculating the initial conformal block derivatives. The idea is to change variables in (\ref{recurse3}) to get a differential equation in terms of $r$ which may be solved by a power series. Since $\Delta$ is a solution of the indicial equation, the coefficients in $g_{\Delta, \ell}^{\Delta_{12}, \Delta_{34}}(r, 1) = r^\Delta \sum_{k = 0}^\infty b_k r^k$ may be determined recursively starting from $b_k = \delta_{k, 0}$. Nothing stops us from applying this method in general dimension but we avoid doing so because it typically leads to polynomials of higher degree.\footnote{It is nevertheless true that this method generates tables more quickly than (\ref{recurse2}) since it omits the $\eta \neq 1$ information. We thank Slava Rychkov for pointing this out.}

The Frobenius method tells us to solve
\begin{equation}
k(\Delta + \ell + k - 1)(\Delta - \ell + k - 2\nu - 1)(2\Delta + k - 2\nu - 2)b_k = \sum_{i = 1}^7 \gamma_{i, k} b_{k - i} \; . \label{recurse5}
\end{equation}
The radial version of the operator (\ref{recurse3}) needed for this step was given in \cite{hor13}. The $\gamma_{i, k}$ were not so we now list them for completeness.
\begin{eqnarray}
\gamma_{1, k} &=& 2c_2(2\nu + 1)(4S + 1) - c_4 + 8P\nu(2\nu + 1) \nonumber \\
&& - 2(\Delta + k - 1)[c_2(2\nu + 1) + 2P(6\nu - 1) + 8S(c_2 + \nu - 2\nu^2)] \nonumber \\
&& + 2(\Delta + k - 2)_2[c_2 + \nu - 2\nu^2 + 4P + 12S(1 - 2\nu)] \nonumber \\
&& + 2(\Delta + k - 3)_3(2\nu - 1 + 8S) - (\Delta + k - 4)_4 \nonumber \\
\gamma_{2, k} &=& 3c_4 - 8P[2S(1 - 6\nu) + 6\nu^2 - 5\nu] + 2c_2[4S(4S + 2\nu + 1) + 2\nu - 3] \nonumber \\
&& - 2(\Delta + k - 2)[2P(16S - 10\nu + 4) + 8S(c_2 + \nu - 2\nu^2) + (1 - 2\nu)(c_2 + 2\nu - 2 + 32S^2)] \nonumber \\
&& - 2(\Delta + k - 3)_2[3c_2 + 7\nu + 2\nu^2 - 10 + 4P + 4S(10S + 6\nu - 3)] \nonumber \\
&& + 2(\Delta + k - 4)_3(7 - 2\nu + 8S) + 3(\Delta + k - 5)_4 \nonumber
\end{eqnarray}

\begin{eqnarray}
\gamma_{3, k} &=& 3c_4 + 2c_2(16S^2 + 2\nu - 3) + 16PS(8S + 2\nu + 5) \nonumber \\
&& - 2(\Delta + k - 3)[(1 - 2\nu)(c_2 + 2\nu - 2 + 32S^2 - 4P) - 8S(8S^2 + 4P + 4\nu^2 + 2c_2 - 5)] \nonumber \\
&& - 2(\Delta + k - 4)_2[8P + 40S^2 + 48S + 3c_2 + 2\nu^2 + 7\nu - 10] \nonumber \\
&& + 2(\Delta + k - 5)_3(7 - 2\nu - 16S) + 3(\Delta + k - 6)_4 \nonumber \\
\gamma_{4, k} &=& 2c_2(3 - 2\nu - 16S^2) - 3c_4 - 16PS(8S + 2\nu + 5) \nonumber \\
&& - 2(\Delta + k - 4) \left [ 12 + 4\nu - 8\nu^2 - c_2(2\nu + 5) \right. \nonumber \\
&& \left. - 8S(8S^2 + 8S\nu + 6S + 4\nu^2 + 2c_2 - 5) + 4P(2\nu - 5 - 8S) \right ] \nonumber \\
&& - 2(\Delta + k - 5)_2[3c_2 + 2\nu^2 + 7\nu - 10 + 8P + S(40S - 18\nu + 21)] \nonumber \\
&& - 2(\Delta + k - 6)_3(16S + 2\nu + 11) - 3(\Delta + k - 7)_4 \label{long-gammas} \\
\gamma_{5, k} &=& 8P(2S(1 - 6\nu) + 6\nu^2 - 5\nu) - 3c_4 - 2c_2[4S(4S + 2\nu + 1) + 2\nu - 3] \nonumber \\
&& - 2(\Delta + k - 5) \left [ 12 + 4\nu - 8\nu^2 - c_2(2\nu + 5 - 8S) \right. \nonumber \\
&& \left. + 2P(3 - 10\nu + 16S) - 8S(2\nu^2 + 5\nu + 3 + 8S\nu + 6S) \right ] \nonumber \\
&& - 2(\Delta + k - 6)_2[22 + 5\nu - 2\nu^2 - 3c_2 - 4P - 4S(10S + 6\nu + 9)] \nonumber \\
&& + 2(\Delta + k - 7)_3(8S - 2\nu - 11) - 3(\Delta + k - 8)_4 \nonumber \\
\gamma_{6, k} &=& c_4 - 2c_2(2\nu + 1)(4S + 1) + 8P\nu(2\nu + 1) \nonumber \\
&& - 2(\Delta + k - 6)[(2\nu + 3)(c_2 - 2\nu - 2) + 6P(2\nu + 1) + 8S(c_2 - 2\nu^2 - 5\nu - 3)] \nonumber \\
&& - 2(\Delta + k - 7)_2[c_2 + 4P - (2\nu + 3)(\nu + 4 + 12S)] \nonumber \\
&& + 2(\Delta + k - 8)_3(2\nu + 5 + 8S) + (\Delta + k - 9)_4 \nonumber \\
\gamma_{7, k} &=& (k + 2\nu - 5)(2\Delta + k - 7)(\Delta + k - \ell - 6)(\Delta + k + \ell + 2\nu - 6) \nonumber
\end{eqnarray}
In order to continue the practice of storing numerators and denominators separately, we use (\ref{recurse5}) to calculate $b_k$ coefficients multiplied by polynomials of degree $3k$. After this, the second last coefficient must be multiplied by three factors that are only in the last one. The third last coefficient must be multiplied by six factors that are only in the last two, \textit{etc}. The case $P = S = 0$ is especially simple because we only have to use $\gamma_{2, k}$, $\gamma_{4, k}$ and $\gamma_{6, k}$ when our conformal block is an expansion in $r^2$. In the rest of this section, we take a closer look at how meromorphic conformal blocks remain finite as $d$ approaches an even integer. This is independent of the rest of the paper since there is no code in \texttt{PyCFTBoot} that makes use of it.\footnote{An especially slow piece of code for this used to be present. Interested readers may find it in the program's \texttt{git} history.}

From (\ref{residues}), we see that only $c_2^{\Delta_{12}, \Delta_{34}}(\ell, k)$ can ever be infinite. This reflects the fact that a pair of coincident poles in Table \ref{data} must always involve series 2. Problematic terms where equal poles are subtracted may cancel in one of two ways:
\begin{enumerate}
\item A term like this that multiplies an expression with $\Delta$ may combine with an infinite residue that multiplies a similar expression with $\Delta$. Consider $\frac{1}{\{\nu\}} \frac{1}{\Delta - \Delta_1} - \frac{1}{\Delta - \Delta_2} \frac{1}{\Delta_2 + n - \Delta_3}$ where we have split $\nu = \left \lfloor \nu \right \rfloor + \{\nu\}$ into its integer and fractional part. If $\Delta_2 = \Delta_1 - \{\nu\}$ and $\Delta_3 = \Delta_2 + n - \{\nu\}$, we may rewrite this as $\frac{1}{\{\nu\}} \left ( \frac{1}{\Delta - \Delta_1} - \frac{1}{\Delta - \Delta_1 + \{\nu\}} \right ) = \frac{1}{(\Delta - \Delta_1)(\Delta - \Delta_1 + \{\nu\})}$ which has a finite limit.
\item The residue being divided by a difference of equal poles might be proportional to $\{\nu\}$ itself.
\end{enumerate}
To see the first type of cancellation, we may set $\Delta_{12}$, $\Delta_{34}$, $\ell$ and $\left \lfloor \nu \right \rfloor$ to zero. In this case $h_{\infty, \ell}(r, 1) = \frac{1}{1 - r^2}$. Going up to $r^4$,
\begin{eqnarray}
h_{\Delta, 0} &=& \frac{1}{1 - r^2} + \frac{c_1(0, 2) r^2}{\Delta + 1} h_{1, 2} + \frac{r^4}{1 - r^2} \left [ \frac{c_1(0, 4)}{\Delta + 3} + \frac{c_2(0, 2)}{\Delta + 1 - \nu} \right ] \nonumber \\
&=& \frac{1}{1 - r^2} + \frac{r^2}{1 - r^2} \frac{c_1(0, 2)}{\Delta + 1} + \frac{r^4}{1 - r^2} \frac{c_1(0, 2)}{\Delta + 1} \left ( \frac{c_1(2, 2)}{4} - \frac{c_3(2, 2)}{2\nu} \right ) \nonumber \\ 
&& + \frac{r^4}{1 - r^2} \left [ \frac{c_1(0, 4)}{\Delta + 3} + \frac{c_2(0, 2)}{\Delta + 1 - \nu} \right ] \; .
\end{eqnarray}
We may now focus on what is proportional to $r^4$. Terms in square brackets come from the first level of the recurrence relation while terms in round brackets come from the second. Taking one of each, we may form the combination
\begin{eqnarray}
\frac{c_2(0, 2)}{\Delta + 1 - \nu} - \frac{c_3(2, 2)}{2\nu} \frac{c_1(0, 2)}{\Delta + 1} &=& \frac{1}{4\nu} \left ( \frac{1}{\Delta + 1 - \nu} - \frac{1}{\Delta + 1} \right ) \nonumber \\
&=& \frac{1}{4(\Delta + 1)(\Delta + 1 - \nu)} \; .
\end{eqnarray}
If all divergences were to cancel in this way, it would make sense to ignore all elements of (\ref{residues}) that are $0$ or $\infty$ and infer their effects later on. For instance, when two poles meant to be subtracted in the denominator coincide, this can be taken as a signal to instead square the pole difference that exists one level up. Unfortunately, because the second type of cancellation is common as well, a robust implementation would have to keep all residues and temporarily equip them with a free symbol for $\{\nu\}$. A 4D recursion to order $r^6$ shows the other phenomenon.
\begin{eqnarray}
h_{\Delta, 0} &=& h_{\infty, 0} + \textcolor{blue}{\frac{r^2 c_1(0, 2)}{\Delta + 1} h_{1, 2}} + \frac{r^2 c_2(0, 1)}{\Delta - 1} h_{3, 0} + \frac{r^4 c_1(0, 4)}{\Delta + 3} h_{1, 4} + \frac{r^4 c_2(0, 2)}{\Delta} h_{4, 0} \nonumber \\
&& + \frac{r^6 c_1(0, 6)}{\Delta + 5} h_{\infty, 6} + \textcolor{red}{\frac{r^6 c_2(0, 3)}{\Delta + 1} h_{\infty, 0}} \label{color1}
\end{eqnarray}
The term in red is infinite and needs to be cancelled by something. This tells us to look at the blue term because it also includes a $\frac{1}{\Delta + 1}$. Expanding this meromorphic block and not setting its dimension to 1 yet,
\begin{eqnarray}
h_{\Delta, 2} &=& h_{\infty, 2} + \frac{r^2 c_1(2, 2)}{\Delta + 3} h_{-1, 4} + \textcolor{magenta}{\frac{r^2 c_2(2, 1)}{\Delta - 1} h_{3, 2}} + \frac{r^2 c_3(2, 2)}{\Delta - 3} h_{5, 0} \nonumber \\
&& + \frac{r^4 c_1(2, 4)}{\Delta + 5} h_{-1, 6} + \frac{r^4 c_2(2, 2)}{\Delta} h_{4, 2} \; . \label{color2}
\end{eqnarray}
The term in magenta cannot be ignored. Even though $c_2(2, 1)$ vanishes with $\{\nu\}$, so does $\Delta - 1$ once we substitute the dimension. Using the fact that this is finite to plug (\ref{color2}) into itself one more time, we see that the $\Delta - 3$ term provides the next divergence. This is what gives the blue term a divergence two levels up allowing it to cancel the red one. Since double poles appear at all levels of the recursion, algebraic simplifications need to be performed repeatedly, slowing down the calculation. Moreover, they only work correctly if all terms are placed over a common denominator --- not just the ones with the free variable $\Delta$. This causes exponentially large numerators and denominators to accumulate during the calculation of $\textbf{d}_{\ell, i}$ even when the fractions themselves are small. Neglecting the error introduced by this would require many more digits than those kept by \cite{eppssv14, s15} and other high precision studies.

\section{Overall structure}
We now describe the three objects in a typical \texttt{PyCFTBoot} session that involve tables of polynomials in $\Delta$. These include an object for the semidefinite program itself which is most directly relevant for the user. The steps described so far ending with convolution amount to two lines of code with mostly self-explanatory arguments.
\begin{eqnarray}
&& \texttt{table1 = ConformalBlockTable(dim, k\_max, l\_max, m\_max, n\_max,} \nonumber \\
&& \texttt{delta\_12, delta\_34, odd\_spins = True)} \nonumber \\
&& \texttt{table2 = ConvolvedBlockTable(table1, symmetric = True)} \nonumber
\end{eqnarray}
A slower version of \texttt{ConformalBlockTable} which should almost never be needed is \\ \texttt{ConformalBlockTableSeed} (or \texttt{ConformalBlockTableSeed2} in even dimension). Although it is meant to be used internally to prepare a (\ref{recurse3}) recursion by calculating the first three derivatives, it can be used to calculate more derivatives explicitly as well. Global variables affecting these two lines of code are \texttt{prec} and \texttt{cutoff}. The default value of \texttt{prec} (the binary precision) is 660, consistent with the 200 decimal digits of \texttt{SDPB}'s example code. The $\theta$ variable in (\ref{pole-approximation}) is \texttt{cutoff} which must be set manually if the user wants it to differ from 0. Another variable which should be set manually is \texttt{sdpb\_path} unless the default value of \texttt{/usr/bin/sdpb} is correct. In addition to these global variables, there are global symbols defined in Table \ref{symbols}.
\begin{table}
\begin{center}
\begin{tabular}{l|l}
Symbol & Description \\
\hline
\texttt{delta} & The scaling dimension variable on which all polynomials depend. \\
\texttt{delta\_ext} & A placeholder for $\Delta_{\phi}$ in \texttt{ConvolvedConformalBlock}. \\
\texttt{ell} & A variable for the few situations that need an unspecified spin.
\end{tabular}
\end{center}
\vspace{-0.5cm}
\caption{These global variable of \texttt{PyCFTBoot} are symbols in the sense that they are treated as variables for the computer algebra.}
\label{symbols}
\end{table}
Two optional parameters have been set to \texttt{True}. For \texttt{odd\_spins}, this indicates that odd spins from $0$ to $\ell_{\mathrm{max}}$ should not be skipped. For \texttt{symmetric}, it indicates that $F_{+, \Delta, \ell}$ is being calculated rather than the default $F_{-, \Delta, \ell}$. There are two other optional parameters that could have been passed above. For \texttt{ConformalBlockTable}, the \texttt{name} parameter tells it to ignore all other arguments, avoid doing any calculation and instead prepare a conformal block table by reading a file. These files are generated by calling \texttt{table1.dump("filename")}. For \texttt{ConvolvedBlockTable}, the \texttt{content} parameter tells the class to produce a linear combination of convolved conformal blocks with prescribed coefficients if the operators are part of a larger (\textit{e.g.} superconformal) multiplet. The elements of this list need further explanation. If one term in the linear combination is a regular convolved block, a subsequent term is specified by three things: an expression for the coefficient, a number indicating how different its $\Delta$ is and an integer indicating how different its $\ell$ is. An artificial example is a multiplet which has conformal blocks (and hence convolved conformal blocks) arranged as follows.
\begin{eqnarray}
\mathcal{G}_{\Delta, \ell} &=& \frac{1}{\Delta + \ell} g_{\Delta, \ell} + \Delta g_{\Delta - 1, \ell + 1} \nonumber \\
\mathcal{F}_{\pm, \Delta, \ell} &=& \frac{1}{\Delta + \ell} F_{\pm, \Delta, \ell} + \Delta F_{\pm, \Delta - 1, \ell + 1} \; \label{fake1}
\end{eqnarray}
In this case, one needs to multiply everything by $\Delta + \ell$ to avoid breakage due to non-polynomial terms. Afterwards, this two element linear combination where each term has three pieces of data, is passed as a pair of triples. One simply gives
\begin{equation}
\texttt{content = [[1, 0, 0], [delta * (delta + ell), -1, 1]]} \nonumber
\end{equation}
to \texttt{ConvolvedBlockTable}.

\subsection{Working with SDPs}
The final class to discuss, the \texttt{SDP}, specifies the arrangement of convolved conformal blocks that needs to vanish for crossing symmetry to hold. The fundamental objects for these sum rules are
\begin{equation}
F_{\pm, \Delta, \ell}^{ij ; kl}(u, v) = v^{\frac{\Delta_j + \Delta_k}{2}} g_{\Delta, \ell}^{\Delta_{ij}, \Delta_{kl}}(u, v) \pm u^{\frac{\Delta_j + \Delta_k}{2}} g_{\Delta, \ell}^{\Delta_{ij}, \Delta_{kl}}(v, u) \; . \label{pre-sdp-block}
\end{equation}
The $\Delta_{\phi}$ variable in (\ref{convolved-block}) may be replaced with all possible values of $\frac{\Delta_j + \Delta_k}{2}$ that can be made from the correlator system under consideration. Linear combinations of the (\ref{pre-sdp-block}) blocks need to give zero in all crossing equations. The weights for these are built out of OPE coefficients which are real by unitarity. When $i = j = k = l$, we simply have squares of OPE coefficients which are positive. In this case, the equation $\sum_{\mathcal{O}} \lambda_{\mathcal{O}}^2 F_{-, \Delta, \ell}(u,v) = 0$ rules out a CFT whenever some functional $\Lambda$ is positive on all $F_{-, \Delta, \ell}$. In more complicated cases, we do not necessarily have positive coefficients. One example \cite{kps14} is the crossing equation with no global symmetry:
\begin{equation}
\sum_{\mathcal{O}} \left [ \lambda_{ij\mathcal{O}} \lambda_{kl\mathcal{O}} F_{\mp, \Delta, \ell}^{ij ; kl}(u, v) \pm \lambda_{kj\mathcal{O}} \lambda_{il\mathcal{O}} F_{\mp, \Delta, \ell}^{kj ; il}(u, v) \right ] = 0 \; . \label{crossing-equation}
\end{equation}
Here, it is not useful to find a $\Lambda$ sending all $F_{\pm, \Delta, \ell}^{ij ; kl}$ to a positive number. What we must do is find a $\Lambda$ that sends particular groupings of them to a positive definite matrix. This is an example of a polynomial matrix problem (a special case of semidefinite program)
\begin{eqnarray}
\mathrm{maximize} && \Lambda \cdot o \nonumber \\
\mathrm{such \; that} && \Lambda \cdot P_{\ell, R}(x) \succeq 0 \;\; \mathrm{for \; all} \;\; x \geq 0, \ell, R \; \label{definition} \\
&& \Lambda \cdot n = 1 \nonumber
\end{eqnarray}
which is what \texttt{SDPB} solves. The objective $o$, the normalization $n$ and the exact relation between $x$ and $\Delta$ are not needed to initialize an \texttt{SDP} class. However, the representations $R$ and the groupings of blocks mentioned above need to be passed in a parameter. Let us call this parameter \texttt{info} and imagine that our correlator system has two operators $\sigma$ and $\epsilon$ with $(\Delta_{\sigma}, \Delta_{\epsilon}) = (0.7, 1.5)$. If \texttt{table3} is another \texttt{ConvolvedBlockTable} instance like \texttt{table2} above, we may call
\begin{equation}
\texttt{sdp = SDP([0.7, 1.5], [table2, table3], vector\_types = info)} \nonumber
\end{equation}
to get a new \texttt{SDP}. The tables and dimensions above may be specified in an arbitrary order but indices describing their positions in the list are obtained from \texttt{info}. The \texttt{vector\_types} argument is required unless both of the first two arguments are single elements. Suppose that the sum in (\ref{crossing-equation}) runs over one representation and all spins. Even and odd spins are considered separately so from the point of view of \texttt{PyCFTBoot}, this leads to two representations $A$ and $B$ which we label with \texttt{0} and \texttt{1} respectively.
\begin{equation}
\texttt{info = [[info1, 2, 0], [info2, 3, 1]]} \nonumber
\end{equation}
The \texttt{2} and \texttt{3} have been chosen because any even integer denotes even spin and any odd integer denotes odd spin. Note that \texttt{info = [[info2, 3, 1], [info1, 2, 0]]} would be wrong because the first representation must be the one containing the identity operator. Now suppose that there are three crossing equations with $2 \times 2$ matrices in the $A$ parts and $1 \times 1$ matrices in the $B$ parts. They might look something like
\begin{equation}
\sum_{\mathcal{O} \in A} \left ( \lambda_{\sigma\sigma\mathcal{O}} \; \lambda_{\epsilon\epsilon\mathcal{O}} \right ) \left [
\begin{tabular}{c}
$\left ( \begin{tabular}{cc} $0$ & $\frac{1}{2} F^{\sigma\sigma ; \sigma\sigma}_{-, \Delta, \ell}$ \\ $\frac{1}{2} F^{\sigma\sigma ; \sigma\sigma}_{-, \Delta, \ell}$ & $0$ \end{tabular} \right )$ \\
$\left ( \begin{tabular}{cc} $F^{\sigma\sigma ; \sigma\sigma}_{-, \Delta, \ell}$ & $0$ \\ $0$ & $F^{\sigma\sigma ; \sigma\sigma}_{-, \Delta, \ell}$ \end{tabular} \right )$ \\
$\left ( \begin{tabular}{cc} $0$ & $\frac{1}{2} F^{\epsilon\epsilon ; \epsilon\epsilon}_{-, \Delta, \ell}$ \\ $\frac{1}{2} F^{\epsilon\epsilon ; \epsilon\epsilon}_{-, \Delta, \ell}$ & $0$ \end{tabular} \right )$
\end{tabular}
\right ] \left (
\begin{tabular}{c}
$\lambda_{\sigma\sigma\mathcal{O}}$ \\ $\lambda_{\epsilon\epsilon\mathcal{O}}$
\end{tabular}
\right ) + \sum_{\mathcal{O} \in B} \lambda^2_{\sigma \epsilon \mathcal{O}} \left [
\begin{tabular}{c}
$0$ \\ $F^{\sigma \epsilon ; \sigma \epsilon}_{+, \Delta, \ell}$ \\ $\frac{3}{2} F^{\sigma \epsilon ; \sigma \epsilon}_{+, \Delta, \ell}$
\end{tabular}
\right ] = 0 \; . \label{fake2}
\end{equation}
Triples of matrices are easy to specify with Python but each matrix element is encoded by four pieces of information: a real coefficient, an integer labelling the convolved conformal block and integers labelling the inner two ($j, k$ in the (\ref{pre-sdp-block}) notation) dimensions. If our \texttt{SDP} is applicable to this system, its \texttt{[table2, table3]} list contains one symmetric convolved block with $\Delta_{\sigma \epsilon}$ differences and one antisymmetric convolved block with $0$ differences. If they appear in this order, the innermost lists of \texttt{info1} have a \texttt{1} in the second position while those of \texttt{info2} have a \texttt{0}. Indeed, one may check that
\begin{eqnarray}
\texttt{info1 =}&\texttt{[[[[0.0, 1, 0, 0], [0.5, 1, 0, 0]], [[0.5, 1, 0, 0], [0.0, 1, 0, 0]]],} \nonumber \\
&\texttt{[[[1.0, 1, 0, 0], [0.0, 1, 0, 0]], [[0.0, 1, 0, 0], [1.0, 1, 0, 0]]],} \nonumber \\
&\texttt{[[[0.0, 1, 0, 0], [0.5, 1, 1, 1]], [[0.5, 1, 1, 1], [0.0, 1, 0, 0]]]]} \nonumber \\
\texttt{info2 =}&\texttt{[[0.0, 0, 0, 0], [1.0, 0, 0, 1], [1.5, 0, 0, 1]]} \nonumber
\end{eqnarray}
fully describes this artificial example.

Before we describe the various ways in which \texttt{SDPB} can be called to do the heavy lifting, it is useful to explore the structure of the allocated \texttt{SDP}. Most of the memory is occupied by \texttt{table}, a three-dimensional list storing the polynomials in $\Delta$. The first index runs over operators from the (\ref{crossing-equation}) sum rule meaning spins and representations. The second and third indices label elements of the matrices that must become positive definite under $\Lambda$. These come in the order given by \texttt{vector\_types}. In (\ref{fake2}), consider two indices \texttt{a} and \texttt{b} on either side of the ``middle'' \texttt{len(sdp.table) / 2} element. Since \texttt{a} corresponds to an $A$ operator, it is perfeclty valid for the user to type \texttt{sdp.table[a][0][1]} or \texttt{sdp.table[a][1][0]}. With $B$ operators however, only \texttt{sdp.table[b][0][0]} is a valid query. Elements thus returned correspond to infinite-dimensional functions, but we have already gone to great lengths to approximate each of these with a finite-dimensional vector of derivatives evaluated at the crossing symmetric point. The object storing this type of truncation is called a \texttt{PolynomialVector}. It has three attributes of which \texttt{vector} is the most important. This is what stores the actual polynomials in $\Delta$. They are essentially the convolved conformal block polynomials from (\ref{pos-poly2}) except they have been multiplied by the appropriate coefficients in \texttt{vector\_types}. The length of something like \texttt{sdp.table[b][0][0].vector} depends on $m_{\mathrm{max}}$ and $n_{\mathrm{max}}$ since each element is a derivative. However, derivatives are often repeated when the sum rule is vectorial. Looking at (\ref{fake2}), instead of simply seeing infinite-dimensional functions, each term is a \textit{triple} of infinite-dimensional functions. \texttt{PyCFTBoot} concatenates the derivatives used to approximate each one. This makes it difficult to remember what each polynomial represents. For instance, if we were naive enough to include no $b$ derivatives, \texttt{sdp.table[b][0][0].vector} would be
\begin{equation}
\left [ 0 \; 0 \; 0 \; P_{+,b}^{\sigma\epsilon ; \sigma\epsilon ; 00}(\Delta) \; P_{+,b}^{\sigma\epsilon ; \sigma\epsilon ; 20}(\Delta) \; P_{+,b}^{\sigma\epsilon ; \sigma\epsilon ; 40}(\Delta) \; \frac{3}{2} P_{+,b}^{\sigma\epsilon ; \sigma\epsilon ; 00}(\Delta) \; \frac{3}{2} P_{+,b}^{\sigma\epsilon ; \sigma\epsilon ; 20}(\Delta) \; \frac{3}{2} P_{+,b}^{\sigma\epsilon ; \sigma\epsilon ; 40}(\Delta) \right ]^{\mathrm{T}} \; .
\end{equation}
The \texttt{SDP} type includes two lists that remind us of where different derivatives are positioned. To see how many $a$ and $b$ derivatives are encoded by a given element, one only needs to check the corresponding elements of \texttt{sdp.m\_order} and \texttt{sdp.n\_order} respectively. The two other attributes of a \texttt{PolynomialVector} --- \texttt{poles} and \texttt{label} --- are also Python lists. The elements of \texttt{poles} are the poles from Table \ref{data} that must be used to reconstruct the positive prefactor of (\ref{pos-poly2}). The \texttt{label} is a two element list with a spin first and a representation label second.\footnote{The \texttt{table} attributes of \texttt{ConformalBlockTable} and \texttt{ConvolvedBlockTable} have very similar layouts. Because no vectors of matrices are present at this stage, no derivatives are repeated in the \texttt{PolynomialVector}s and only one index is needed to iterate over them. Inspecting their \texttt{label} attributes, we see that the second element is always \texttt{0}. This is because no other labels have been given in \texttt{vector\_types} yet.} One more interesting attribute is \texttt{sdp.unit}, the contribution of the identity. This is the one operator that is guaranteed to appear in every crossing equation. To calculate this, \texttt{SDP} substitutes $\Delta = \ell = 0$ into the \texttt{table} elements that have $R$ as the singlet representation. It also multiplies by the proper OPE coefficients. These are known because the canonically normalized
\begin{eqnarray}
\left < \phi_i(x_1) \phi_j(x_2) \right > &=& \frac{\delta_{ij}}{|x_{12}|^{\Delta_i + \Delta_j}} \nonumber \\
\left < \phi_i(x_1) \phi_j(x_2) \phi_k(x_3) \right > &=& \frac{\lambda_{ijk}}{|x_{12}|^{\Delta_i + \Delta_j - \Delta_k}|x_{23}|^{\Delta_j + \Delta_k - \Delta_i}|x_{13}|^{\Delta_k + \Delta_i - \Delta_j}}
\end{eqnarray}
are only consistent with each other if all $\lambda_{ijI} = 1$.

When numerically excluding CFTs, the most obvious physical inputs are the allowed ranges for the scaling dimensions in a trial spectrum. When $\Delta \in [\Delta_{\mathrm{min}}, \infty)$, all polynomials should have $\Delta$ replaced by $\Delta_{\mathrm{min}} + x$ so that $x$ satisfies the positivity requirement in (\ref{definition}). In a CFT that is unitary but otherwise unconstrained, $\Delta_{\mathrm{min}}$ is equal to the unitarity bound,
\begin{equation}
\Delta_{\mathrm{unitary}} = \begin{cases} \frac{d - 2}{2} & \ell = 0 \\ d + \ell - 2 & \ell > 0 \end{cases} \; . \label{unitarity}
\end{equation}
Although bounds for the \texttt{SDP} class are always initialized to (\ref{unitarity}), they may be changed with a call to \texttt{sdp.set\_bound([l, r], delta\_min)}. Here \texttt{l} is a spin and \texttt{r} is a representation label. These bounds continue to be enforced until they are undone manually. Resetting a given $(\ell, R)$ to the unitarity bound is done with \texttt{sdp.set\_bound([l, r])}. Omitting both arguments causes \texttt{PyCFTBoot} to reset the bounds of all operators. In the exact same manner, individual points may be added with \texttt{sdp.add\_point([l, r], delta\_value)}. These are explicitly allowed dimensions at which \texttt{PolynomialVector}s should be evaluated. Calling \texttt{sdp.add\_point([0, 0], 1.0)} prepares us for bootstrapping a theory with spin-0 singlets of dimension 1, even after something like \texttt{sdp.set\_bound([0, 0], 1.2)} has been called. Again, removing points for a given operator type or all operator types may be accomplished by omitting arguments. The last persistently stored property of an \texttt{SDP} is the list of options passed to \texttt{SDPB}. The options that \texttt{PyCFTBoot} correctly chooses without user interaction are \texttt{--precision} and all options not passed as key-value pairs. For everything else, a helper function is provided. As an example, one may leave some processor resources unused by passing the key-value pair \texttt{--maxThreads=2}. \texttt{PyCFTBoot} can be told to use this with the method \texttt{sdp.set\_option("maxThreads", 2)}. The line undoing this is \texttt{sdp.set\_option("maxThreads")} and the line undoing everything is \texttt{sdp.set\_option()}.

\subsection{Writing XML files}
\texttt{SDPB} learns everything that it needs to know about an optimization from an XML file \cite{s15}. Knowing that the points and bounds determine $x$ in (\ref{definition}) while \texttt{sdp.table} determines the polynomials, only the objective $o$ and the normalization $n$ are needed to write the XML. To rule out CFTs with a certain gap, one chooses an objective vector of zero and a normalization of \texttt{sdp.unit}. Another common task is maximizing a squared OPE coefficient. For this, the objective vector must be \texttt{sdp.unit} with the \texttt{PolynomialVector} for the $(\Delta, \ell, R)$ being maximized as the normalization. These are specified using \texttt{sdp.write\_xml(obj, norm, "name")} but it is often not necessary to call this function directly. A more convenient function is an implementation of the bisection described in \cite{rrtv08}, which works for identical scalars. To bisect over gaps in an $(\ell, R)$ operator, one should call \texttt{sdp.bisect(lower, upper, tol, [l, r])}. Since this method finds upper bounds, \texttt{upper} should be a gap where a $\Lambda$ solving (\ref{definition}) exists and \texttt{lower} should be a gap where such a $\Lambda$ does not. The boundary between allowed and disallowed regions is returned with a tolerance of \texttt{tol}. Since points on either side of the boundary take different amounts of time to test, it is convenient to use a biased binary search. Once a long test taking time $L$ and a short test taking time $S$ have been performed, we may approximate the optimal bias $\beta$ by minimizing the time:
\begin{eqnarray}
T &=& \#(\beta) [\beta L + (1 - \beta)S] \nonumber \\
&\propto& \frac{\beta L + (1 - \beta)S}{\log \left [ \beta^\beta (1 - \beta)^{(1 - \beta)} \right ]} \; .
\end{eqnarray}
We have used the fact that $\beta$ is also the probability that a given test will reduce our uncertainty by a factor of $\beta$. Strictly speaking, when \texttt{SDPB} finishes finding a functional, the problem is called primal-dual optimal (another word for ``primal and dual feasible''). The bisection in \texttt{PyCFTBoot} does not wait for this to happen. Rather, it takes advantage of a very safe assumption: when dual feasibility (time $L$) is achieved before primal feasibility (time $S$) during \texttt{SDPB}'s iterations, it is only a matter of time before primal feasibility is achieved as well.\footnote{One must be careful when assuming the converse: that only unsolvable problems will achieve primal feasibility first. Sometimes when testing a point far from the boundary, a primal-dual optimal solution will be approached with the opposite order.} If the full solution functional is desired after a bisection, it may be found with \texttt{sdp.solution\_functional(gap, [l, r])}. For an excluded $\Delta$, the returned functional must turn $(\ell, R)$'s matrix of \texttt{PolynomialVector}s into something positive definite. It is therefore useful to tune $\Delta$ until the determinant of this matrix is exactly zero. These zeros, returned by \texttt{sdp.extremal\_dimensions(functional, [l, r])}, are exactly the scaling dimensions in the spectrum of a CFT that lives on the boundary \cite{ep13}. Rather than obeying $\Lambda \cdot n = 1$, the functionals used by these methods are normalized to have a leading component of $1$. This reflects the alternate definition of a semidefinite program used by \texttt{SDPB}. Instead of (\ref{definition}), the program solves
\begin{eqnarray}
\mathrm{maximize} && \Lambda \cdot \tilde{o} \nonumber \\
\mathrm{such \; that} && \Lambda \cdot \tilde{P}_{\ell, R}(x) \succeq 0 \;\; \mathrm{for \; all} \;\; x \geq 0, \ell, R \; \label{definition-alt} \\
&& \Lambda_0 = 1 \nonumber
\end{eqnarray}
which is trivially equivalent \cite{s15}. One simply substitutes $\Lambda_0 = \frac{1}{n_0} \left ( 1 - \sum_{i = 1}^N \Lambda_i n_i  \right )$ into (\ref{definition}). When we once again collect all terms proportional to $\Lambda_i$, we find that the $i^{\mathrm{th}}$ component of the reshuffled $\tilde{P}_{\ell, R}$ involves components $i$ and $0$ of $P_{\ell, R}$. Finally, we should describe \texttt{PyCFTBoot}'s built-in method for bounding OPE coefficients. The logic for this is easiest to write in the single correlator case: $\sum_{\mathcal{O} \neq I} \lambda^2_{\mathcal{O}} F_{-, \mathcal{O}} = -F_{-, I}$. Normalizing $\Lambda$ on the convolved conformal block of some particular $\mathcal{O}^{\prime}$ \cite{cr09},
\begin{equation}
\lambda_{\mathcal{O}^{\prime}}^2 = \Lambda(F_{-, I}) - \sum_{\mathcal{O} \neq I, \mathcal{O}^{\prime}} \lambda^2_{\mathcal{O}} \Lambda(F_{-, \mathcal{O}}) \leq \Lambda(F_{-, I}) \; . \label{opemax}
\end{equation}
The last step of neglecting the strictly positive terms is still valid in the multi-correlator case. However, if multiple OPE coefficients in the crossing equations involve the operator $\mathcal{O}^{\prime}$, the left hand side of (\ref{opemax}) will involve the quadratic form $\bm{\lambda}_{\mathcal{O}^{\prime}}^{\mathrm{T}} \Lambda(M_{\mathcal{O}^{\prime}}) \bm{\lambda}_{\mathcal{O}^{\prime}}$. In this case, only the length of the OPE coefficient vector will be bounded. Since $\Lambda$ turns the vector of matrices $M_{\mathcal{O}^{\prime}}$ into a single positive definite matrix, we simply use the fact that the inequality is preserved if we replace this matrix with the identity multiplied by its smallest eigenvalue. This technique of using $\Lambda(F_{-, I})$ to bound a squared OPE coefficient is implemented by \texttt{sdp.opemax(delta\_value, [l, r])}. Here, $(\Delta, \ell, R)$ are the quantum numbers of $\mathcal{O}^{\prime}$.

Users may notice a time delay when allocating an \texttt{SDP} class. This is used for a calculation involving the positive $\chi_{\ell}(\Delta)$ prefactors in (\ref{pos-poly2}) which have been largely ignored up to this point. Even though the semidefinite program itself is not affected, incorporating these into the XML file can significantly improve the performance and numerical stability of \texttt{SDPB} \cite{s15}. The non-trivial step is the calculation of a bilinear basis --- a set of polynomials that are orthogonal with respect to the $\chi_{\ell}(\Delta_{\mathrm{min}} + x)$ measure on $(0, \infty)$. Since these functions only change when the bounds change, the bases do not need to be recalculated every time an XML file is written. After \texttt{PyCFTBoot} allocates an \texttt{SDP} and calculates all bases, a particular $\ell$'s basis is only recalculated when \texttt{sdp.set\_bound([l, r])} is called. This saves time during a bisection because the many XML files generated only have different bounds for a single operator type. Multi-correlator bootstraps, which cannot use bisection, typically have all of their XML files generated by different instances of \texttt{SDP}. To avoid a performance hit in this case, \texttt{PyCFTBoot} provides an optional argument to \texttt{SDP} called \texttt{prototype}. This allows an existing \texttt{SDP} to have its bilinear basis recycled in the allocation of a new one. For completeness, we now review the most direct method for finding $m + 1$ polynomials orthogonal under
\begin{equation}
\chi_{\ell}(x) = \frac{r_*^{x + \Delta_{\mathrm{min}}(\ell)}}{\prod_i \left ( x + \Delta_{\mathrm{min}}(\ell) - \Delta_i(\ell) \right )} \; .
\end{equation}
We may clearly multiply by $r_*^{\Delta_{\mathrm{min}}}$ and shift the poles so we will omit $\Delta_{\mathrm{min}}$ in what follows. What we are trying to find is the matrix
\begin{equation}
L = \left [
\begin{tabular}{ccccc}
$q_{00}$ & $0$ & $0$ & $\dots$ & $0$ \\
$q_{10}$ & $q_{11}$ & $0$ & $\dots$ & $0$ \\
$\vdots$ & $\vdots$ & $\vdots$ & $\ddots$ & $\vdots$ \\
$q_{m0}$ & $q_{m1}$ & $q_{m2}$ & $\dots$ & $q_{mm}$
\end{tabular}
\right ] \; ,
\end{equation}
where the $i^{\mathrm{th}}$ polynomial is $q_i(x) = q_{i0} + q_{i1}x + \dots + q_{ii}x^i$. The statement of orthonormality is
\begin{equation}
\int_0^{\infty} L \left [ \begin{tabular}{c} $1$ \\ $\vdots$ \\ $x^m$ \end{tabular} \right ] [ 1 \; \dots \; x^m ] L^{\mathrm{T}} \chi(x) \textup{d}x = I \; . \label{orthonormality}
\end{equation}
If we let $M_{ij} = \left < x^i, x^j \right >$ elements come from the positive definite matrix of inner products, (\ref{orthonormality}) says that $LML^{\mathrm{T}} = I$. Since $M = L^{-1} L^{-1 \mathrm{T}}$, $L$ is the inverse of the lower triangular matrix in the Cholesky decomposition of $M$ \cite{t12}. $M$ has anti-diagonal bands because $M_{ij}$ is fully determined by $i + j$. Therefore, our problem reduces to the evaluation of $2m + 1$
\begin{eqnarray}
\int_0^{\infty} \frac{x^n r_*^x}{\prod_i (x - \Delta_i)} \textup{d}x &=& \sum_i \frac{1}{\prod_{j \neq i} (\Delta_j - \Delta_i)} \int_0^{\infty} \frac{x^n r_*^x}{x - \Delta_i} \textup{d}x \nonumber \\
&=& \sum_i \frac{1}{(-\log r_*)^n \prod_{j \neq i} (\Delta_j - \Delta_i)} \int_0^{\infty} \frac{y^n e^{-y}}{y + \Delta_i \log r_*} \textup{d}y \label{1pole-integral}
\end{eqnarray}
integrals. Above, we have used a partial fraction decomposition valid for simple poles and made the substitution $y = -x \log r_*$. Allowing for incomplete gamma functions, the integral in (\ref{1pole-integral}) may be evaluated explicitly. This is because the integral representation
\begin{equation}
\Gamma(-n, z) = \frac{z^{-n} e^{-z}}{\Gamma(1 + n)} \int_0^{\infty} \frac{y^n e^{-y}}{y + z} \textup{d}y
\end{equation}
follows from the relation between the upper incomplete gamma function and the ${}_1F_1$ hypergeometric function. A related expression which may arise in even $d$ is
\begin{equation}
\int_0^{\infty} \frac{y^n e^{-y}}{(y + z)^2} \textup{d}y = nz^{n - 1} e^z \Gamma(n) \left [ \Gamma(1 - n, z) - z\Gamma(-n, z) \right ] \; .
\end{equation}
This follows from partial integration. By handling the slightly more complicated partial fraction decomposition, which \texttt{PyCFTBoot} and the \texttt{SDPB} example code can easily do, this allows us to still find $M$ when $\chi_{\ell}$ has double poles. We should note that the Cholesky decomposition of $M$ can only be found when all of its eigenvalues are significantly greater than 0. Even if the desired polynomial precision is fairly low, this requirement forces us to keep many digits before the polynomials can be found at all. It would be interesting to see if indirect polynomial algorithms \cite{g94} can make this step less demanding.

\section{Some examples}
This section contains longer code snippets to show how the main tasks in the conformal bootstrap can be accomplished. It can also be used for reference since examples are often preferable to more verbose documentation. \texttt{PyCFTBoot} is four files at the time of writing but importing \texttt{bootstrap.py} will automatically import the others. They should be placed in the working directory or one of the system directories searched by Python.

\subsection{Identical scalars}
We begin with the simplest bound: the dimension of $\phi^2$ in terms of the dimension of $\phi$. There must be a $\mathbb{Z}_2$ symmetry for these to be different. To set this up in \texttt{PyCFTBoot}, we run:
\begin{eqnarray}
&& \texttt{>>> from bootstrap import *} \nonumber \\
&& \texttt{>>> table1 = ConformalBlockTable(3, 15, 15, 1, 3)} \nonumber \\
&& \texttt{>>> table2 = ConvolvedBlockTable(table1)} \nonumber
\end{eqnarray}
Recall that this sets up a $d = 3$ table with 15 poles, 15 spins and a $1 \times 3$ triangle of derivatives. If we are interested in the $\Delta_{\phi^2}$ close to where the Ising model is known to live, we should run:
\begin{eqnarray}
&& \texttt{>>> sdp = SDP(0.52, table2)} \nonumber \\
&& \texttt{>>> result = sdp.bisect(0.7, 1.7, 0.01, 0)} \nonumber \\
&& \texttt{>>> result} \nonumber \\
&& \texttt{1.434375} \nonumber
\end{eqnarray}
Although we are limiting ourselves to an error of $1\%$, the error is more likely dominated by missing derivatives. The last argument of \texttt{0} (\texttt{[0, 0]} is only necessary when there is more global symmetry) states that we are bounding scalars in the $\phi \times \phi$ OPE rather than operators of higher spin.

\subsection{A faster approach}
This bound may be bisected more quickly if we use \cite{kps13}'s method to only increase the polynomial degree for sufficiently large residues. Since the \texttt{cutoff} controlling this is a global variable, it can only be altered if we keep it in its own namespace.
\begin{eqnarray}
&& \texttt{>>> import bootstrap} \nonumber \\
&& \texttt{>>> bootstrap.cutoff = 1e-10} \nonumber \\
&& \texttt{>>> table1 = bootstrap.ConformalBlockTable(3, 15, 15, 1, 3)} \nonumber \\
&& \texttt{>>> table2 = bootstrap.ConvolvedBlockTable(table1)} \nonumber \\
&& \texttt{>>> sdp = bootstrap.SDP(0.52, table2)} \nonumber \\
&& \texttt{>>> result = sdp.bisect(0.7, 1.7, 0.01, 0)} \nonumber \\
&& \texttt{>>> result} \nonumber \\
&& \texttt{1.434375} \nonumber
\end{eqnarray}
The output from \texttt{SDPB} (supressed above) shows that the iterations take less time and that they are slightly fewer in number.

\subsection{The next scalar}
The results above state that consistent CFTs at $\Delta_{\phi} = 0.52$ stop existing once the scalar part of the search space starts at $1.44$. The search space is allowed to be both discrete and continuous, so even if it starts at $1.434375$, we may still require that gaps for all other operators are significantly greater. Beginning in the same way as before,
\begin{eqnarray}
&& \texttt{>>> sdp = SDP(0.52, table2)} \nonumber \\
&& \texttt{>>> sdp.add\_point(0, 1.434375)} \nonumber \\
&& \texttt{>>> result = sdp.bisect(1.44, 8.0, 0.01, 0)} \nonumber \\
&& \texttt{>>> result} \nonumber \\
&& \texttt{4.8225} \nonumber
\end{eqnarray}
where we have taken the extra step of adding a point. This tells us that one possible CFT living near the edge of the $\Delta_{\phi^2}$ bound has $4.8225$ as the dimension of its next $\mathbb{Z}_2$-even scalar. However, spectra saturating these bounds are unique \cite{ep13}. The extremal functional method exploits this fact to find several low-lying dimensions at once without having to repeat the above procedure. To use this in \texttt{PyCFTBoot}, we need to set $\Delta_{\phi^2}$ to something slightly larger than $1.434375$ where crossing symmetry holds. By our bisection, the closest value where an extremal functional exists is at most $0.01$ more than this.
\begin{eqnarray}
&& \texttt{>>> sdp = SDP(0.52, table2)} \nonumber \\
&& \texttt{>>> func = sdp.solution\_functional(1.434375 + 0.01, 0)} \nonumber \\
&& \texttt{>>> spec = sdp.extremal\_dimensions(func, 0)} \nonumber \\
&& \texttt{>>> spec} \nonumber \\
&& \texttt{[0.95491336461809961, 1.4442910577901338, 4.3268882750844018]} \nonumber
\end{eqnarray}
The \texttt{0} arguments above are again short for $(\ell, R)$ labels of \texttt{[0, 0]}. The three dimensions returned in the spectrum include an inadmissible value, the bound we imposed and the desired second scalar. This time, it appears much closer to the high precision estimate obtained in \cite{eppssv14}.

\subsection{Imposing another gap}
The previous example shows us how to fix $\phi^2$ and then constrain higher scalars in $\phi \times \phi$. It is often just as useful to proceed in the other direction. For example, the allowed region in $(\Delta_{\phi}, \Delta_{\phi^2})$ space shrinks if we first make additional assumptions on the second $\mathbb{Z}_2$-even scalar. This is particularly natural in the Ising model, which constrains this operator to be irrelevant by definition. We need to increase the derivative order and kept pole order to see a noticeable effect:
\begin{eqnarray}
&& \texttt{>>> import bootstrap} \nonumber \\
&& \texttt{>>> bootstrap.cutoff = 1e-10} \nonumber \\
&& \texttt{>>> sig = 0.52} \nonumber \\
&& \texttt{>>> eps = 1.42} \nonumber \\
&& \texttt{>>> table1 = bootstrap.ConformalBlockTable(3, 20, 15, 2, 4)} \nonumber \\
&& \texttt{>>> table2 = bootstrap.ConvolvedBlockTable(table1)} \nonumber
\end{eqnarray}
It only makes sense to bisect when the boundary of the allowed region is the graph of some function. This is no longer the case when we demand that the dimensions of two internal operators are a certain distance apart.
\begin{eqnarray}
&& \texttt{>>> sdp = bootstrap.SDP(sig, table2)} \nonumber \\
&& \texttt{>>> sdp.set\_bound(0, 3.0)} \nonumber \\
&& \texttt{>>> sdp.add\_point(0, eps)} \nonumber \\
&& \texttt{>>> result = sdp.iterate()} \nonumber \\
&& \texttt{>>> result} \nonumber \\
&& \texttt{True} \nonumber
\end{eqnarray}
Clearly a point close to the Ising model is still allowed. However, if we start with \texttt{(sig, eps) = (0.52, 1.2)} and keep the rest of the code the same, it can easily be seen that \texttt{sdp.iterate()} returns \texttt{False} and begins to reveal a non-trivial shape. A repeated scan over many different values is what produces the plot in \cite{eppssv12} with a sharp corner.

\subsection{OPE maximization}
Non-trivial features also appear in OPE coefficient bounds. As explained in (\ref{opemax}), \texttt{PyCFTBoot} has a method for dealing with this part of the CFT data as well. Using the same setup as before,
\begin{eqnarray}
&& \texttt{>>> sdp = bootstrap.SDP(0.52, table2)} \nonumber \\
&& \texttt{>>> result1 = sdp.opemax(3.0, 2)} \nonumber \\
&& \texttt{>>> float(result1)} \nonumber \\
&& \texttt{18.76458372724582} \nonumber
\end{eqnarray}
Here, we have chosen to maximize the coefficient of a spin-2 operator evaluated $\Delta_{\mathrm{unitary}} = 3$. In other words, this bounds $\lambda_T^2$, the coefficient of the stress-energy tensor. The returned value is careful to account for the positive prefactors $\chi_\ell(\Delta)$ but necessarily still depends on our normalization convention for confromal blocks. The best way to extract the physical information is to compare this to another OPE coefficient. Below, we do this at the (almost) free field theory point.
\begin{eqnarray}
&& \texttt{>>> sdp = bootstrap.SDP(0.5001, table2)} \nonumber \\
&& \texttt{>>> result2 = sdp.opemax(3.0, 2)} \nonumber \\
&& \texttt{>>> float(result2)} \nonumber \\
&& \texttt{16.14053403869564} \nonumber
\end{eqnarray}
Dividing one result by the other causes any extra factors to cancel out. Because we are dealing with the stress-energy tensor, we have enough information to find the central charge
\begin{equation}
C_T = \frac{d}{d - 1} \left ( \frac{\Delta_{\phi}}{\lambda_T} \right )^2 \; .
\end{equation}
Checking that the one from \texttt{result1} is about $93\%$ of the one from \texttt{result2}, we have verified the results of \cite{eppssv12, eppssv14} which studied the central charge in the vicinity of the Ising point.

\subsection{Global symmetry}
The examples so far have all treated a single crossing equation. One way to go beyond this is to give our external scalars flavour indices under some global Lie group symmetry. Fundamentals of $SO(N)$ provide a simple yet important example. Their OPE may be written schematically as
\begin{equation}
\phi_i \times \phi_j \sim \sum_{\mathcal{O} \in S} \delta_{ij} \mathcal{O} + \sum_{\mathcal{O} \in T} \mathcal{O}_{\{i, j\}} + \sum_{\mathcal{O} \in A} \mathcal{O}_{[i, j]} \; , \label{global-ope}
\end{equation}
in terms of singlet, traceless symmetric and antisymmetric tensor structures. Terms with Fermi symmetry may only couple to odd spins just as Bose symmetry allowed us to only consider even spins before. Keeping all spins in the tables we prepare, we also need to perform symmetric and antisymmetric convolutions.
\begin{eqnarray}
&& \texttt{>>> from bootstrap import *} \nonumber \\
&& \texttt{>>> table1 = ConformalBlockTable(3, 15, 15, 1, 3, odd\_spins = True)} \nonumber \\
&& \texttt{>>> table2 = ConvolvedBlockTable(table1, symmetric = True)} \nonumber \\
&& \texttt{>>> table3 = ConvolvedBlockTable(table1)} \nonumber
\end{eqnarray}
Now that our tables include odd spins, our sign convention for conformal blocks becomes especially important. Being careful with this, the sum rule that follows from (\ref{global-ope}) is
\begin{equation}
\sum_{\mathcal{O} \in S} \lambda^2_{\mathcal{O}} \left [ \begin{tabular}{c} $0$ \\ $F_{-, \Delta, \ell}$ \\ $F_{+, \Delta, \ell}$ \end{tabular} \right ]
+ \sum_{\mathcal{O} \in T} \lambda^2_{\mathcal{O}} \left [ \begin{tabular}{c} $F_{-, \Delta, \ell}$ \\ $\left ( 1 - \frac{2}{N} \right ) F_{-, \Delta, \ell}$ \\ $- \left ( 1 + \frac{2}{N} \right ) F_{+, \Delta, \ell}$ \end{tabular} \right ]
+ \sum_{\mathcal{O} \in A} \lambda^2_{\mathcal{O}} \left [ \begin{tabular}{c} $F_{-, \Delta, \ell}$ \\ $-F_{-, \Delta, \ell}$ \\ $F_{+, \Delta, \ell}$ \end{tabular} \right ] = 0 \; . \label{global-rule}
\end{equation}
The last vector differs from what appears in \cite{kps13} by a sign. The presence of $(-1)^{\ell}$ in (\ref{leading-block}) is what tells us to introduce this sign when using \texttt{PyCFTBoot}. It can be seen in \cite{kpsv15} that the same authors have now switched to the normalization used here. Because the dimensions of the $\phi_i$ are all the same, entries of these vectors may be specified with two numbers instead of four. It is easy to do this after choosing an $N$ and a list of tables.
\begin{eqnarray}
&& \texttt{>>> N = 3.0} \nonumber \\
&& \texttt{>>> table\_list = [table2, table3]} \nonumber \\
&& \texttt{>>> vec1 = [[0, 1], [1, 1], [1, 0]]} \nonumber \\
&& \texttt{>>> vec2 = [[1, 1], [1.0 - (2.0 / N), 1], [-(1.0 + (2.0 / N)), 0]]} \nonumber \\
&& \texttt{>>> vec3 = [[1, 1], [-1, 1], [1, 0]]} \nonumber
\end{eqnarray}
To formulate (\ref{global-rule}) we just need to give these vectors (even, even, odd) spins and (\texttt{0}, \texttt{1}, \texttt{2}) representation labels.
\begin{eqnarray}
&& \texttt{>>> info = [[vec1, 0, 0], [vec2, 0, 1], [vec3, 1, 2]]} \nonumber \\
&& \texttt{>>> sdp = SDP(0.52, table\_list, vector\_types = info)} \nonumber \\
&& \texttt{>>> result = sdp.bisect(0.7, 1.8, 0.01, [0, 0])} \nonumber \\
&& \texttt{>>> result} \nonumber \\
&& \texttt{1.6453125000000002} \nonumber
\end{eqnarray}
This value is approximately what it should be, looking at the bound on singlet scalars produced in \cite{kps13}.

\subsection{Mixed correlators}
Applying the above methods when four point functions have arbitrary scaling dimensions represents an important advance for the bootstrap. This can reveal a wealth of information even for the simplest CFTs because OPEs with no dimension difference are blind to $\mathbb{Z}_2$-odd operators. Following \cite{kps14} where many more details can be found, we present an example with only $\mathbb{Z}_2$ symmetry. The even $\epsilon \in E$ is now more than just an internal operator being summed over. It is a member of the four point function on the same level as the odd $\sigma \in O$. Letting the indices in (\ref{crossing-equation}) run over all combinations of $\sigma$ and $\epsilon$, we derive a number of crossing equations that can be put into matrix form.
\begin{equation}
\sum_{\mathcal{O} \in E, 2 | \ell} \left ( \lambda_{\sigma\sigma\mathcal{O}} \; \lambda_{\epsilon\epsilon\mathcal{O}} \right ) V_{E, \Delta, \ell} \left (
\begin{tabular}{c}
$\lambda_{\sigma\sigma\mathcal{O}}$ \\ $\lambda_{\epsilon\epsilon\mathcal{O}}$
\end{tabular}
\right ) + \sum_{\mathcal{O} \in O, 2 | \ell} \lambda^2_{\sigma \epsilon \mathcal{O}} V_{O+, \Delta, \ell}
+ \sum_{\mathcal{O} \in O, 2 \nmid \ell} \lambda^2_{\sigma \epsilon \mathcal{O}} V_{O-, \Delta, \ell} = 0 \label{mixed-rule}
\end{equation}
where
\begin{equation}
V_{E, \Delta, \ell} = \left [
\begin{tabular}{c}
$\left ( \begin{tabular}{cc} $F_{-, \Delta, \ell}^{\sigma \sigma ; \sigma \sigma}$ & $0$ \\ $0$ & $0$ \end{tabular} \right )$ \\
$\left ( \begin{tabular}{cc} $0$ & $0$ \\ $0$ & $F_{-, \Delta, \ell}^{\epsilon \epsilon ; \epsilon \epsilon}$ \end{tabular} \right )$ \\
$\left ( \begin{tabular}{cc} $0$ & $0$ \\ $0$ & $0$ \end{tabular} \right )$ \\
$\left ( \begin{tabular}{cc} $0$ & $\frac{1}{2} F_{-, \Delta, \ell}^{\sigma \sigma ; \epsilon \epsilon}$ \\ $\frac{1}{2} F_{-, \Delta, \ell}^{\sigma \sigma ; \epsilon \epsilon}$ & $0$ \end{tabular} \right )$ \\
$\left ( \begin{tabular}{cc} $0$ & $\frac{1}{2} F_{+, \Delta, \ell}^{\sigma \sigma ; \epsilon \epsilon}$ \\ $\frac{1}{2} F_{+, \Delta, \ell}^{\sigma \sigma ; \epsilon \epsilon}$ & $0$ \end{tabular} \right )$
\end{tabular}
\right ] \; , \; 
V_{O+, \Delta, \ell} = \left [ \begin{tabular}{c} $0$ \\ $0$ \\ $F_{-, \Delta, \ell}^{\sigma \epsilon ; \sigma \epsilon}$ \\ $F_{-, \Delta, \ell}^{\epsilon \sigma ; \sigma \epsilon}$ \\ $-F_{+, \Delta, \ell}^{\epsilon \sigma ; \sigma \epsilon}$ \end{tabular} \right ] \; , \;
V_{O-, \Delta, \ell} = \left [ \begin{tabular}{c} $0$ \\ $0$ \\ $F_{-, \Delta, \ell}^{\sigma \epsilon ; \sigma \epsilon}$ \\ $-F_{-, \Delta, \ell}^{\epsilon \sigma ; \sigma \epsilon}$ \\ $F_{+, \Delta, \ell}^{\epsilon \sigma ; \sigma \epsilon}$ \end{tabular} \right ] \; .
\end{equation}
The only difference with respect to \cite{kps14} is our choice not to use factors of $(-1)^{\ell}$ to combine the $O$ sums over even and odd spins in (\ref{mixed-rule}). Looking at all possible dimension differences, there are three conformal block tables to make which give rise to five convolutions.
\begin{eqnarray}
&& \texttt{>>> from bootstrap import *} \nonumber \\
&& \texttt{>>> sig = 0.518} \nonumber \\
&& \texttt{>>> eps = 1.412} \nonumber \\
&& \texttt{>>> g\_tab1 = ConformalBlockTable(3, 20, 20, 2, 4)} \nonumber \\
&& \texttt{>>> g\_tab2 = ConformalBlockTable(3, 20, 20, 2, 4, \textbackslash} \nonumber \\
&& \texttt{... eps - sig, sig - eps, odd\_spins = True)} \nonumber \\
&& \texttt{>>> g\_tab3 = ConformalBlockTable(3, 20, 20, 2, 4, \textbackslash} \nonumber \\
&& \texttt{... sig - eps, sig - eps, odd\_spins = True)} \nonumber \\
&& \texttt{>>> f\_tab1a = ConvolvedBlockTable(g\_tab1)} \nonumber \\
&& \texttt{>>> f\_tab1s = ConvolvedBlockTable(g\_tab1, symmetric = True)} \nonumber \\
&& \texttt{>>> f\_tab2a = ConvolvedBlockTable(g\_tab2)} \nonumber \\
&& \texttt{>>> f\_tab2s = ConvolvedBlockTable(g\_tab2, symmetric = True)} \nonumber \\
&& \texttt{>>> f\_tab3 = ConvolvedBlockTable(g\_tab3)} \nonumber \\
&& \texttt{>>> dim\_list = [sig, eps]} \nonumber \\
&& \texttt{>>> tab\_list = [f\_tab1a, f\_tab1s, f\_tab2a, f\_tab2s, f\_tab3]} \nonumber
\end{eqnarray}
Entering $V_{O\pm, \Delta, \ell}$ is similar to our syntax for the vectors in (\ref{global-rule}). However, we need two \texttt{dim\_list} indices after the \texttt{tab\_list} index to specify the inner $\sigma$s and $\epsilon$s.
\begin{eqnarray}
\texttt{>>> v2 = [[0, 0, 0, 0], [0, 0, 0, 0], [1, 4, 1, 0], [1, 2, 0, 0], [-1, 3, 0, 0]]} \nonumber \\
\texttt{>>> v3 = [[0, 0, 0, 0], [0, 0, 0, 0], [1, 4, 1, 0], [-1, 2, 0, 0], [1, 3, 0, 0]]} \nonumber
\end{eqnarray}
To continue with $V_{E, \Delta, \ell}$, each entry should be a $2 \times 2$ matrix in the standard Python notation.
\begin{eqnarray}
&& \texttt{>>> m1 = [[[1, 0, 0, 0], [0, 0, 0, 0]], [[0, 0, 0, 0], [0, 0, 0, 0]]]} \nonumber \\
&& \texttt{>>> m2 = [[[0, 0, 0, 0], [0, 0, 0, 0]], [[0, 0, 0, 0], [1, 0, 1, 1]]]} \nonumber \\
&& \texttt{>>> m3 = [[[0, 0, 0, 0], [0, 0, 0, 0]], [[0, 0, 0, 0], [0, 0, 0, 0]]]} \nonumber \\
&& \texttt{>>> m4 = [[[0, 0, 0, 0], [0.5, 0, 0, 1]], [[0.5, 0, 0, 1], [0, 0, 0, 0]]]} \nonumber \\
&& \texttt{>>> m5 = [[[0, 1, 0, 0], [0.5, 1, 0, 1]], [[0.5, 1, 0, 1], [0, 1, 0, 0]]]} \nonumber \\
&& \texttt{>>> v1 = [m1, m2, m3, m4, m5]} \nonumber
\end{eqnarray}
Folding these into the final argument of \texttt{SDP} and iterating can now be done in a familiar way. We should also use an option to ensure that anything taking much longer than a primal feasible problem is correctly recognized as a dual feasible problem.
\begin{eqnarray}
&& \texttt{>>> info = [[v1, 0, 0], [v2, 0, 1], [v3, 1, 2]]} \nonumber \\
&& \texttt{>>> sdp = SDP(dim\_list, tab\_list, vector\_types = info)} \nonumber \\
&& \texttt{>>> sdp.set\_option("dualErrorThreshold", 1e-15)} \nonumber \\
&& \texttt{>>> sdp.add\_point([0, 2], sig)} \nonumber \\
&& \texttt{>>> sdp.set\_bound([0, 2], 3.0)} \nonumber \\
&& \texttt{>>> sdp.set\_bound([0, 0], eps)} \nonumber \\
&& \texttt{>>> result = sdp.iterate()} \nonumber \\
&& \texttt{>>> result} \nonumber \\
&& \texttt{True} \nonumber
\end{eqnarray}
The only $\mathbb{Z}_2$-even bound we have set is the one that defines $\epsilon$. Instead, the power of this bound comes from our $\mathbb{Z}_2$-odd statement, that every such operator except $\sigma$ has $\Delta > 3$. Repeating this example with \texttt{(sig, eps) = (0.518, 1.2)} returns \texttt{False} which is exactly what we saw with the $\mathbb{Z}_2$-even gap before. However, this time we can also rule out CFTs by going ``right'' of the Ising model in $(\Delta_{\sigma}, \Delta_{\epsilon})$ space. Since \texttt{(sig, eps) = (0.53, 1.412)} also returns \texttt{False}, we begin to see hints that we are exploring an isolated region of allowed scaling dimensions.

\subsection{A superconformal example}
Among the known constructions of conformal field theories, examples without supersymmetry are relatively rare. In adapting out CFT bootstrap methods to handle SCFTs, the main new step is combining conformal blocks for different operators in the same multiplet. For the example of 4D $\mathcal{N} = 1$ chiral primaries, \cite{ps10} found the following blocks adding to the results of \cite{do02}:
\begin{eqnarray}
\mathcal{G}_{\Delta, \ell} &=& g_{\Delta, \ell} - \frac{(\ell + 2)(\Delta + \ell)}{(\ell + 1)(\Delta + \ell + 1)} g_{\Delta + 1, \ell + 1} - \frac{\ell (\Delta - \ell - 2)}{(\ell + 1)(\Delta - \ell - 1)} g_{\Delta + 1, \ell - 1} \nonumber \\
&& + \frac{(\Delta + \ell)(\Delta - \ell - 2)}{(\Delta + \ell + 1)(\Delta - \ell - 1)} g_{\Delta + 2, \ell} \; . \label{super-block1}
\end{eqnarray}
The R-symmetry in this case is $U(1) \simeq SO(2)$ which allows us to write
\begin{equation}
\sum_{\mathcal{O} \in S, 2 | \ell} \lambda^2_{\mathcal{O}} \left [ \begin{tabular}{c} $F_{-, \Delta, \ell}$ \\ $F_{-, \Delta, \ell}$ \\ $F_{+, \Delta, \ell}$ \end{tabular} \right ]
+ \sum_{\mathcal{O} \in T, 2 | \ell} \lambda^2_{\mathcal{O}} \left [ \begin{tabular}{c} $0$ \\ $2F_{-, \Delta, \ell}$ \\ $-2F_{+, \Delta, \ell}$ \end{tabular} \right ]
+ \sum_{\mathcal{O} \in S, 2 \nmid \ell} \lambda^2_{\mathcal{O}} \left [ \begin{tabular}{c} $-F_{-, \Delta, \ell}$ \\ $F_{-, \Delta, \ell}$ \\ $F_{+, \Delta, \ell}$ \end{tabular} \right ] = 0 \; . \label{super-rule1}
\end{equation}
One change compared to (\ref{global-rule}) is that we have recognized antisymmetric $A$ operators as simply being odd-spin singlets. The other is that we have replaced the middle row with itself plus twice the top row and replaced the top row with the middle row. Due to the $(-1)^{\ell}$ factor in (\ref{leading-block}), we have paid attention to the middle two terms of (\ref{super-block1}) and the last term of (\ref{super-rule1}). Otherwise the normalization of our $\mathcal{N} = 1$ block is the same as that of \cite{cgikpy15}. Seeing the difference between the first two rows above, it becomes clear that a final sum rule will also have to involve
\begin{eqnarray}
\tilde{\mathcal{G}}_{\Delta, \ell} &=& g_{\Delta, \ell} + \frac{(\ell + 2)(\Delta + \ell)}{(\ell + 1)(\Delta + \ell + 1)} g_{\Delta + 1, \ell + 1} + \frac{\ell (\Delta - \ell - 2)}{(\ell + 1)(\Delta - \ell - 1)} g_{\Delta + 1, \ell - 1} \nonumber \\
&& + \frac{(\Delta + \ell)(\Delta - \ell - 2)}{(\Delta + \ell + 1)(\Delta - \ell - 1)} g_{\Delta + 2, \ell} \; . \label{super-block2}
\end{eqnarray}
Lines that encode this in \texttt{PyCFTBoot} are
\begin{eqnarray}
&& \texttt{>>> import bootstrap} \nonumber \\
&& \texttt{>>> c1 = (delta + ell + 1) * (delta - ell - 1) * (ell + 1)} \nonumber \\
&& \texttt{>>> c2 = -(delta + ell) * (delta - ell - 1) * (ell + 2)} \nonumber \\
&& \texttt{>>> c3 = -(delta - ell - 2) * (delta + ell + 1) * ell} \nonumber \\
&& \texttt{>>> c4 = (delta + ell) * (delta - ell - 2) * (ell + 1)} \nonumber \\
&& \texttt{>>> combo1 = [[c1, 0, 0], [c2, 1, 1], [c3, 1, -1], [c4, 2, 0]]} \nonumber \\
&& \texttt{>>> combo2 = combo1} \nonumber \\
&& \texttt{>>> combo2[1][0] *= -1} \nonumber \\
&& \texttt{>>> combo2[2][0] *= -1} \nonumber
\end{eqnarray}
where each triple has a coefficient, a shift in $\Delta$ and then a shift in $\ell$. Uncharged operators in $S$ come from OPEs of the form $\Phi \times \Phi^{\dagger}$. These are the ones that have three other operators related by supersymmetry. There are also the $T$ operators from $\Phi \times \Phi$ OPEs which only make use of regular conformal blocks. Allocating all of the tables we need, it is advisable to keep many derivatives because the convergence of 4D $\mathcal{N} = 1$ bounds is notoriously slow.
\begin{eqnarray}
&& \texttt{>>> g\_tab = ConformalBlockTable(3.99, 25, 26, 3, 5, odd\_spins = True)} \nonumber \\
&& \texttt{>>> f\_tab1a = ConvolvedBlockTable(g\_tab)} \nonumber \\
&& \texttt{>>> f\_tab1s = ConvolvedBlockTable(g\_tab, symmetric = True)} \nonumber \\
&& \texttt{>>> f\_tab2a = ConvolvedBlockTable(g\_tab, content = combo1)} \nonumber \\
&& \texttt{>>> f\_tab2s = ConvolvedBlockTable(g\_tab, symmetric = True, content = combo1)} \nonumber \\
&& \texttt{>>> f\_tab3 = ConvolvedBlockTable(g\_tab, content = combo2)} \nonumber \\
&& \texttt{>>> tab\_list = [f\_tab1a, f\_tab1s, f\_tab2a, f\_tab2s, f\_tab3]} \nonumber
\end{eqnarray}
With 26 spins kept above, the spins of our singlet operators will go up to 25 because each superconformal block draws from the spin above it. The normalization in (\ref{super-block1}) is convenient because the spin-0 expression gives a vanishing coefficient to its $\ell - 1$ term. If this were not the case, \texttt{PyCFTBoot} would still skip any terms telling us to naively include negative spin. The main remaining task is to write (\ref{super-rule1}) in terms of superconformal blocks and absorb a factor of $2$ for convenience.
\begin{equation}
\sum_{\mathcal{O} \in S, 2 | \ell} \lambda^2_{\mathcal{O}} \left [ \begin{tabular}{c} $\tilde{\mathcal{F}}_{-, \Delta, \ell}$ \\ $\mathcal{F}_{-, \Delta, \ell}$ \\ $\mathcal{F}_{+, \Delta, \ell}$ \end{tabular} \right ]
+ \sum_{\mathcal{O} \in S, 2 \nmid \ell} \lambda^2_{\mathcal{O}} \left [ \begin{tabular}{c} $-\tilde{\mathcal{F}}_{-, \Delta, \ell}$ \\ $\mathcal{F}_{-, \Delta, \ell}$ \\ $\mathcal{F}_{+, \Delta, \ell}$ \end{tabular} \right ]
+ \sum_{\mathcal{O} \in T, 2 | \ell} \lambda^2_{\mathcal{O}} \left [ \begin{tabular}{c} $0$ \\ $F_{-, \Delta, \ell}$ \\ $-F_{+, \Delta, \ell}$ \end{tabular} \right ] = 0 \label{super-rule2}
\end{equation}
We will now enter this as an \texttt{SDP} at an external dimension of $\Delta_{\phi} = 1.4$.
\begin{eqnarray}
&& \texttt{>>> vec1 = [[1, 4], [1, 2], [1, 3]]} \nonumber \\
&& \texttt{>>> vec2 = [[-1, 4], [1, 2], [1, 3]]} \nonumber \\
&& \texttt{>>> vec3 = [[0, 0], [1, 0], [-1, 1]]} \nonumber \\
&& \texttt{>>> info = [[vec1, 0, 0], [vec2, 1, 1], [vec3, 0, 2]]} \nonumber \\
&& \texttt{>>> sdp = SDP(1.4, tab\_list, vector\_types = info)} \nonumber
\end{eqnarray}
Since $T$ is constrained by more than just unitarity, we need to set the bound $\Delta_{\mathrm{min}} = |2\Delta_{\phi} - 3| + 3 + \ell$. The only dimensions lower than this in the charged sector belong to BPS operators with $\Delta = 2\Delta_{\phi} + \ell$. Finishing the computation of this bound,
\begin{eqnarray}
&& \texttt{>>> sdp.set\_option("dualErrorThreshold", 1e-22)} \nonumber \\
&& \texttt{>>> for l in range(0, 27, 2):} \nonumber \\
&& \texttt{...     sdp.set\_bound([l, 2], abs(2 * 1.4 - 3) + 3 + l)} \nonumber \\
&& \texttt{...     sdp.add\_point([l, 2], 2 * 1.4 + l)} \nonumber \\
&& \texttt{...} \nonumber \\
&& \texttt{>>> result = sdp.bisect(3.0, 6.0, 0.01, [0, 0])} \nonumber \\
&& \texttt{>>> result} \nonumber \\
&& \texttt{3.966796875} \nonumber
\end{eqnarray}
This is still about $20\%$ away from the known value where a special theory is conjectured to live. The properties of this kink have been studied extensively in \cite{ps15}.

\section{A longer example}
As a final demonstration, we use \texttt{PyCFTBoot} to investigate operator dimensions on the line of critical theories interpolating between the Ising model in $d = 3$ and the free boson in $d = 4$. The most standard argument for these theories comes from the perturbative analysis of Wilson and Fisher \cite{wf72}. Perturbation theory yields insight because $\phi^4$ theory, which is infrared free in $4$ dimensions, has a weakly coupled IR fixed point when the dimension is analytically continued to $4 - \varepsilon$. Since they come from a theory with $\mathbb{Z}_2$ symmetry, all of these fixed points should be in the Ising model's universality class. The operators $\phi$ and $\phi^2$ become what we have been calling $\sigma$ and $\epsilon$ --- the scalar of lowest dimension and the first scalar appearing in the simplest OPE. Making only a unitarity assumption, \cite{eppssv13} used the bootstrap to go beyond perturbation theory and place upper bounds on $\Delta_{\epsilon}$ in terms of $\Delta_{\sigma}$ over the whole range of dimensions. Our goal is to constrain $(\Delta_{\sigma}, \Delta_{\epsilon})$ space further by using the same assumptions that have been successful with the 3D Ising model \cite{kps14, s15}. Namely, we demand that only a single $\mathbb{Z}_2$-odd scalar has a dimension below $d$.

\begin{figure}[h]
\centering
\includegraphics[scale=0.85]{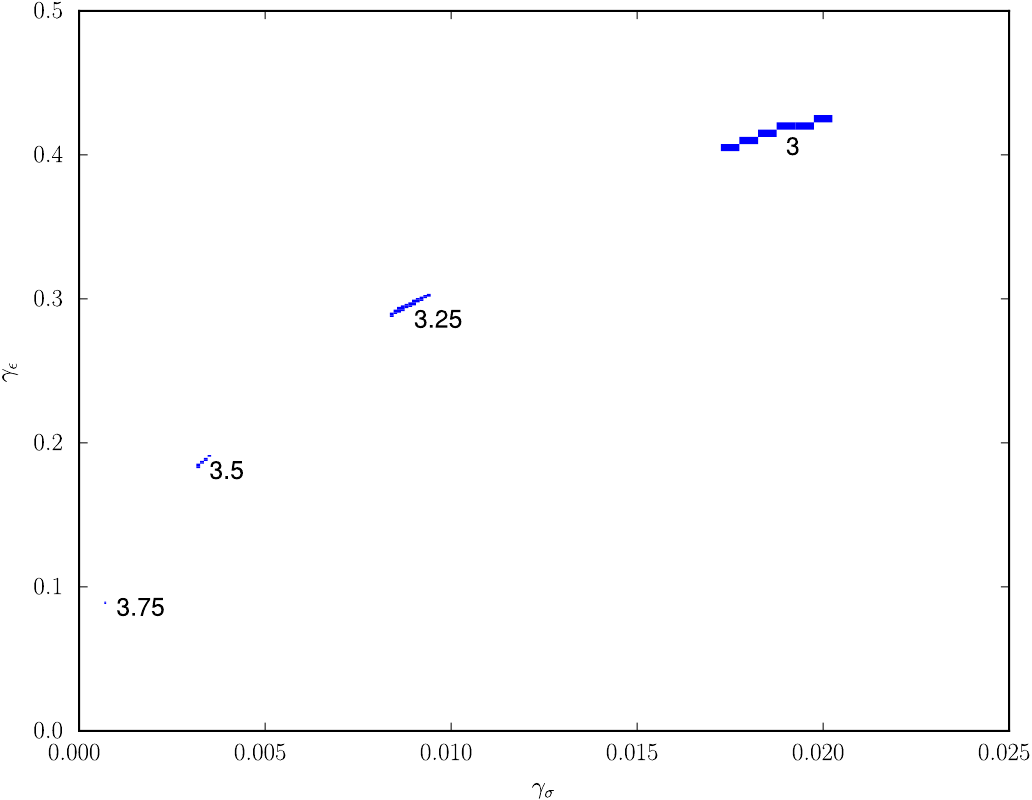}
\caption{Plot of allowed anomalous dimensions in $d \in \{ 3, 3.25, 3.5, 3.75 \}$. For each $d$, there is a closed region of points that cannot be excluded using the constraints of the conformal bootstrap on the correlators $\left < \sigma \sigma \sigma \sigma \right >$, $\left < \sigma \sigma \epsilon \epsilon \right >$, $\left < \epsilon \epsilon \epsilon \epsilon \right >$. Their positions have good agreement with the dimensions of the Wilson-Fisher fixed points calculated with the $\varepsilon$-expansion.}
\label{islands}
\end{figure}

Since only this gap and the conformal blocks depend on $d$, we use the same crossing equations as (\ref{mixed-rule}). For our truncation parameters, we choose $k_{\mathrm{max}} = 30$, $\ell_{\mathrm{max}} = 20$, $m_{\mathrm{max}} = 3$ and $n_{\mathrm{max}} = 5$. The only non-default parameters of \texttt{SDPB} are \texttt{--precision=660 --dualErrorThreshold=1e-15}. Allowed CFTs return \texttt{found primal feasible solution} well before this. Due to the number of points that must be checked, it does not make sense to call \texttt{ConformalBlockTable} every time we get to a new point. It is common for two different points in the region being scanned to have the same dimension differences. For this reason, we use \texttt{PyCFTBoot} to dump all of the tables beforehand and then setup \texttt{SDP}s as a second step. Each \texttt{SDP} is allocated with the \texttt{prototype = old\_sdp} argument since almost all parts of the bilinear basis are shared. The resolutions chosen for our scans are given in Table \ref{resolutions}. When a pair of anomalous dimensions is found to be compatible with crossing symmetry, a pixel of the appropriate width and height is drawn. Anomalous dimensions are defined via
\begin{eqnarray}
\gamma_{\sigma} &=& \Delta_{\sigma} - \frac{d - 2}{2} \nonumber \\
\gamma_{\epsilon} &=& \Delta_{\epsilon} - (d - 2) \; .
\end{eqnarray}
Results of these scans are shown in Figure \ref{islands}. The well known 3D island occupies the top right corner. In the bottom left corner is a barely visible island for $d = 3.75$. It consists of just three points centred at $(\Delta_{\sigma}, \Delta_{\epsilon}) \approx (0.8757, 1.839)$. Although it is likely that larger regions of non-excluded points exist away from each island, we have not attempted to search for them.

\begin{table}[h]
\begin{center}
\begin{tabular}{l|l|l}
$d$ & Horizontal step & Vertical step \\
\hline
$3$ & $0.0005$ & $0.005$ \\
$3.25$ & $0.0001$ & $0.001$ \\
$3.5$ & $0.0001$ & $0.001$ \\
$3.75$ & $0.00005$ & $0.0005$
\end{tabular}
\end{center}
\vspace{-0.5cm}
\caption{Horizontal and vertical spacing between points checked for primal / dual feasibility.}
\label{resolutions}
\end{table}

Numerical checks indicate that the islands shrink as we increase the dimension of our search space. However, their characteristic sizes for a fixed number of derivatives clearly depend on $d$. Whereas the error bar on $\Delta_{\sigma}$ from our 3D Ising scan is about $0.005$, the same computational resources put toward the Wilson-Fisher fixed point in $d = 3.75$ give us an error bar that is two orders of magnitude smaller. It is in fact comparable in size to the second smallest island found for the 3D Ising model in \cite{s15}. This makes sense because increasing $d$ brings us closer to the perturbative regime where error bars from a numerical technique like the bootstrap are not needed at all. This trend is also the reason why we have not decreased $d$ further. Below $3$, the islands continue to grow until they merge with the unbounded regions. In the extreme case of $d = 2$, the same plot that follows purely from crossing symmetry and unitarity is returned with the extra assumptions here having no effect.\footnote{We thank Anton de la Fuente for pointing this out.}

\section{Discussion}
The last example, intended merely to demonstrate the capabilities of \texttt{PyCFTBoot}, was done with much less CPU time than bootstrap calculations designed to break precision records. Indeed, this has not been accomplished. Using Borel summation and agreement with 2D values as a boundary condition, \cite{lz87} has calculated Wilson-Fisher critical exponents up to fifth order in $\varepsilon = 4 - d$. Their values for $d = 3.75$ converted to conformal scaling dimensions are
\begin{eqnarray}
\Delta_{\sigma} &=& 0.875718 \pm 0.000005 \nonumber \\
\Delta_{\epsilon} &=& 1.83943 \pm 0.00005 \; .
\end{eqnarray}
Our bootstrap result, which is in complete agreement, does not fix as many digits. Moreover, it is not necessarily true that our error bars, found by rigorously excluding points are safer to use than those found by resummation techniques. The caveat that prevents this interpretation is the tacit assumption that the Wilson-Fisher fixed point is unitary. While this assumption has long been made, it is incorrect according to a recent analysis which takes non-integer $d$ seriously \cite{hrv15}. A striking result is their finding that four descendants in the spectrum have the complex dimensions
\begin{eqnarray}
\Delta &=& 23 + \left ( \frac{\lambda}{36} - \frac{7}{2} \right ) \varepsilon + O(\varepsilon^2) \nonumber \\
\lambda &\in& \{ 16.93372103 \pm 5.59469106i, 42.88540243 \pm 1.07557547i \} \; .
\end{eqnarray}
These evanescent operators, as they are called, have correlation functions with more familiar operators like $\phi$ and $\phi^2$ that only vanish when $d$ is an integer. The primaries from which they arise are guaranteed to have $\Delta \geq 15$ \cite{hrv15}.

Unlike some more approximate schemes \cite{g13}, the methods used in \texttt{SDPB} and similar codes cannot rule out trial spectra containing complex dimensions. Complex dimensions would motivate us to consider polynomials in $x + iy$ instead of just $x$. Since dimensions occur in conjugate pairs, the only natural domain for $y$ would be all of $\mathbb{R}$. This is incompatible with the positivity conditions of (\ref{definition}) which have to be phrased on a half-line.\footnote{When a polynomial has odd degree for example, it can never be positive for all $x \in \mathbb{R}$. Neglecting odd degrees does not make sense because a conformal block approximation can never become worse when its degree increases by one.} Even if this problem could be solved, adapting the method to non-unitary theories would also require a way of dealing with complex OPE coefficients. It is important then to discuss why Figure \ref{islands} still appears to show four reasonable islands.

For $d$ sufficiently close to $3$, it is obvious that an island will still be found. The bootstrap, like any well-posed computational problem, is robust to small changes in the parameters. These parameters include the spatial $d$ and also the dimensionality of Lie group symmetries that are present \cite{cpy14, cipy15}. On the other hand, there have already been situations where $d$ is fractional enough for the bootstrap to rule out all unitary theories \cite{gp14}. Parameters that impact our ability to make this distinction are the number of poles and the number of derivatives. There are three possibilities for what happens as they are increased:
\begin{enumerate}
\item The islands disappear when these parameters reach certain large but finite numbers.
\item The islands persist because some undiscovered unitary CFT has a low-lying spectrum very similar to that of the Wilson-Fisher fixed point.
\item The islands persist for some other reason.
\end{enumerate}
We conjecture that the first option is realized. In particular, one's ability to find a fake ``Wilson-Fisher island'' with arbitrarily high precision would cast doubt on the conventional wisdom for what happens to the Ising model's island in $d = 3$. A mixed correlator bootstrap that lacks the power to rule out some crossing asymmetric points in $3 < d < 4$, would probably also cause the 3D island to converge to a finite size. While this is a possible topic for future work, it is also likely to be a difficult one. The island for $d = 3.75$ is only small compared to the scale of Figure \ref{islands}. We have no reason to believe that it is anywhere close to disappearing.

Apart from this, there are still many unitary theories that can benefit from a conformal bootstrap treatment. \texttt{PyCFTBoot} can allow these studies to happen more quickly and serve as a starting point for those wishing to test modifications to the various algorithms. Adding code to deal with more general conformal blocks is an important next step. Constraints from external tensor and spinor operators are expected to shed light on a number of previously unexplored theories in $2 < d < 6$. They could also help answer the still open question of whether interacting CFTs above six dimensions can exist.

The list of example problems that \texttt{PyCFTBoot} can handle is already fairly large. We expect this to grow as the community makes progress on important phenomenological questions at an increasing rate. If bugs are encountered along the way, anyone can read the code of \texttt{PyCFTBoot} or one of its dependencies in order to suggest a fix. A few flagship results of the bootstrap have become widely known and the pool of introductory papers is of course larger than it has ever been. Now that adequate software is available for bootstrapping a CFT from start to finish, the time is ripe to get new people involved.

\section*{Acknowledgements}
This work was partially supported by the Natural Sciences and Engineering Research Council of Canada. The numerics were performed on the Pheno cluster in the C. N. Yang Institute for Theoretical Physics. I thank Chi Ming Hung for maintaining this cluster and installing some of the required libraries. Some discussions related to this work also happened during the Simons Summer Workshop 2015. Participants including Agnese Bissi, Alejandro Castedo, Matthijs Hogervorst, Balt van Rees and Alessandro Vichi gave helpful comments while this code was in development. I am especially grateful to Madalena Lemos for sharing advice on all aspects of conformal bootstrap calculations over the course of a year. In the final stages of this work, I received input from Sheer El-Showk, Anton de la Fuente, Leonardo Rastelli, Slava Rychkov and David Simmons-Duffin. After the first release, Junchen Rong found a misprint and Jaehoon Lee helped to debug a crucial normalization error.

\bibliographystyle{unsrt}
\bibliography{references}

\end{document}